\documentclass[journal]{IEEEtran}

\usepackage{cite}
\ifCLASSINFOpdf
\else

\fi

\ifCLASSOPTIONcompsoc
  \usepackage[nocompress]{cite}
  \usepackage{graphicx}
  \usepackage{amsmath}
  \usepackage{xcolor}
  \usepackage{mathtools}
  \usepackage{algorithm}
  \usepackage[noend]{algpseudocode}
  \usepackage{adjustbox}
  \usepackage{amsfonts}
  \usepackage{amssymb}
  \usepackage{multirow}
  \usepackage{float}
  \usepackage{caption}
  \usepackage[bookmarks=true,breaklinks=true,colorlinks,linkcolor=black,citecolor=blue,urlcolor=black]{hyperref}
\else
  \usepackage{cite}
  \usepackage{caption}
  \usepackage{color}
  \usepackage{amsmath}
\fi

\usepackage{tabularx}
\usepackage{graphicx}
\usepackage{subfigure}
\usepackage{amsmath, amsthm, amssymb, amsfonts}
\usepackage{booktabs}    
\makeatletter
\usepackage{tikz}
\usepackage{makecell}
\usepackage{comment}

\usepackage{stfloats}

\usepackage{graphicx}

\begin{document}

\title{Superconductor Logic Implementation with All-JJ Inductor-Free Cell Library}

\author{Haolin Cong, Sasan Razmkhah, Mustafa Altay Karamuftuoglu, Massoud Pedram% <-this % stops a space
\thanks{H. Cong, S. Razmkhah and M. Pedram are with the Department of Electrical and Computer Engineering, University of Southern California, Los Angeles, CA 90007 USA (e-mail: haolinco@usc.edu; razmkhah@usc.edu; karamuft@usc.edu; pedram@usc.edu)}% <-this % stops a space
}

% The paper headers
\markboth{IEEE TRANSACTION ON APPLIED SUPERCONDUCTIVITY}%
{Shell \MakeLowercase{\textit{et al.}}: overlined Demo of IEEEtran.cls for IEEE Journals}

% make the title area
\maketitle

% As a general rule, do not put math, special symbols or citations
% in the abstract or keywords.
\begin{abstract}
\noindent
Single flux quantum (SFQ) technology has garnered significant attention due to its low switching power and high operational speed. Researchers have been actively pursuing more advanced devices and technologies to further reduce the reliance on inductors, bias, and dynamic power. Recently, innovative magnetic Josephson junction devices have emerged, enhancing the field of superconductor electronics (SCE) logic. This paper introduces a novel cell library design that relies entirely on Josephson junctions (JJs), showing promising potential for eliminating the need for inductors in conventional SFQ cells. This results in a 55\% reduction in cell size and an 80\% decrease in both static and dynamic power consumption. The proposed library implements a half flux quantum (HFQ) logic, where each pulse duration is half that of a single flux quantum pulse. The paper presents the schematics of the basic cells, emphasizing critical circuit parameters and their margins. Additionally, it examines layout blueprints, showcasing the advantageous area-saving characteristics of the proposed design.
\end{abstract}

% Note that keywords are not normally used for peer-reviewed papers.
\begin{IEEEkeywords}
Superconductor Electronics, Single Flux Quantum, Half Flux Quantum, All-JJ
\end{IEEEkeywords}

\IEEEpeerreviewmaketitle

\section{INTRODUCTION}

\IEEEPARstart{S}{ingle} flux quantum (SFQ) technology~\cite{Likharev1991} holds great potential for advancing the next generation of VLSI (large-scale integration) circuits. Among SFQ logic circuits, the Rapid SFQ (RSFQ) logic family stands out, offering extremely high operational speeds. RSFQ utilizes Josephson Junctions (JJs), known for their ultra-fast picosecond (ps) switching times. RSFQ logic cells exhibit swift responses with a clock-to-output delay of $10 \mathrm{ps}$, enabling RSFQ systems to excel in the $40 \mathrm{GHz}$ to $60 \mathrm{GHz}$ range~\cite{RSFQ_Kato,RSFQ_Nagaoka,razmkhahBook2023}. Josephson Junctions require significantly less energy to switch than CMOS technology, potentially as low as $10^{-19} \mathrm{J/bit}$. This efficiency sets RSFQ systems apart as more power-efficient alternatives. The RSFQ domain has garnered substantial attention, with studies covering circuit and system designs~\cite{RSFQ_block_Hironaka,RSFQ_system_Kawaguchi,RSFQ_block_Cong}, innovative layout designs~\cite{RSFQ_layout_Fourie,RSFQ_layout_Herbst}, and notable progress in electronic design automation (EDA) tools and algorithms~\cite{RSFQ_EDA_Yang,RSFQ_EDA_Zhang2020}.

Despite the numerous advantages offered by RSFQ circuits, they are confronted with significant challenges. RSFQ circuit integration density remains relatively low, accommodating only around 10,000 logic gates within a $1 \mathrm{cm^2}$ chip area. This limitation is the key limiting factor to meet the logic circuit requirements of various applications. Additionally, RSFQ circuits lack dense on-chip memory and require substantial bias currents for their operation. Given these obstacles, it becomes imperative to explore alternative logic circuit families capable of overcoming these challenges and driving the advancement of SFQ technology.

At the core of superconductor circuits lies the JJ. Serving as the active component in an SFQ circuit, the JJ adopts the standardized Superconductor-Insulator-Superconductor (SIS) configuration. The dynamic behavior of a JJ is encapsulated by the current-phase relation (CPR), which relies on parameters such as the current density ($J_s$), the critical current density ($J_{C}$) at which the JJ exits the superconducting state, and the phase difference ($\phi$) spanning two superconducting layers. This simplified CPR equation assumes a consistent supercurrent tunneling through the JJ's barrier while maintaining temperatures well below the critical threshold. Importantly, this equation accurately approximates the behavior of the SIS JJs and serves as the cornerstone for most SPICE-based simulation engines.

The MITLL SFQ5ee process \cite{MITLL_Tolpygo2015,MITLL_Tolpygo2016} is an exemplary technology that employs a $Nb/Al-AlO_x/Nb$ type junctions, with the superconducting layers composed of $Nb$ and the insulator being $AlO_{x}$. By replacing the insulator layer with a magnetic material featuring a built-in magnetic field, the SIS JJ transforms into a magnetic junction (MJJ) \cite{MJJ_Ryazanov,MJJ_Baek}. The MJJ has been extensively studied for its unique properties, giving rise to innovative devices like the $\pi$-junction ($\pi$-JJ), $\phi$-junction ($\phi$-JJ), and $2\phi$-junction ($2\phi$-JJ), which have piqued the interest of researchers.

Researchers are actively pursuing the development of faster, more efficient, and compact technology. However, reducing the size of inductors in SFQ circuits poses an important challenge due to increasing mutual inductance and cross-talk in SFQ circuit layouts, limiting further reductions in the width and spacing of metal lines. While the kinetic inductor offers a potential solution, it has its own challenges. Addressing this issue, the work outlined in \cite{2phi_design_Soloviev} attempts to eliminate the need for inductors and enhance SFQ circuit scalability by utilizing logic cells (such as AND and OR gates, Not and XOR gates, non-destructive readout D-flipflops) with $2\phi$-junctions. Simultaneously, efforts are underway to leverage the $2\phi$-junction to minimize dynamic power consumption in SFQ systems, as explored in \cite{2phi_design_Salameh}. This study introduced three novel cells incorporating the $2\phi$-junction: a Josephson transmission line (JTL), an inverter, and an OR gate. Compared to conventional RSFQ cells, these cells utilize half flux quantum (HFQ) pulses, resulting in reduced latency and switching power. However, it is important to note that this study lacks detailed circuit parameters, and the three cells do not constitute a complete standard cell library to meet fundamental functional requirements. In another study, an interface is introduced to bridge SFQ circuits and HFQ circuits \cite{RSFQ_half_Hasegawa}.

This paper introduces a comprehensive standard cell library based on $2\phi$-junction technology, encompassing four essential logic cells (inverter, AND gate, OR gate, and XOR gate), five transmission blocks (Josephson transmission line (JTL), splitter, merger, passive transmission line (PTL) transmitter and receiver), one storage cell (DFF), and two I/O interface blocks (DC/SFQ and SFQ/DC converters.) This library caters to the fundamental requirements of a general-use system.

We validate the functionality of each cell using the JoSIM simulator \cite{coldfluxteamSuperTools2023}. Furthermore, we optimize the circuits using qCS \cite{qCSgithub2023}, achieving commendable margins. Critical circuit parameters and their margins are meticulously presented, along with projected layout areas for the cells using the MITLL SFQ5ee process. Notably, a simulated $2\phi$-junction device is included. In Section \ref{sec:standard_cell}, we delve into the design intricacies of each cell within the library, concluding with a summary. Additionally, this section outlines the methodology employed for estimating layouts and offers a comparative analysis with a conventional RSFQ cell library. Lastly, Section \ref{sec:conc} provides a concluding perspective for this paper.

\section{$2\phi$-Junction}
\label{sec:2phi_jj}
\noindent
Recent research has revealed intriguing phenomena at the $0-\pi$ transition, where the fundamental sinusoidal term of the current-phase relation (CPR) vanishes, rendering high-order harmonic terms significant \cite{phi_JJ_Goldobin}. 
In a study by \cite{2phi_Stoutimore}, a single superconductor–ferromagnet–superconductor (SFS) junction employing a $\mathrm{Cu_{47}Ni_{53}}$ alloy barrier is realized with two parallel superconducting inductors: a readout inductor and a small shunt inductor. The readout inductor couples with a commercial DC superconducting quantum interference device (SQUID) sensor, detecting flux $\Phi$ within the readout loop.

Through measurements of the CPR across various barrier thicknesses and temperatures, reference \cite{2phi_Stoutimore} establishes a $\pi$-periodic behavior, putting to rest alternative explanations except for a second-order CPR. Consequently, the CPR is redefined as follows:
\begin{equation} \label{eq:2phi}
J_s(\phi) = J_{c1}\sin(\phi)+J_{c2}\sin(2\phi)
\end{equation}
This gives rise to a new device known as the $2\phi$-JJ, characterized by the CPR:
\begin{equation} \label{eq:2phi_jj}
J_s(\phi) = J_{c2}\sin(2\phi)
\end{equation}
The $2\phi$-JJ possesses intriguing properties: it features a current-phase relation with a period of $\pi$ rather than $2\pi$, undergoes switching with a $\pi$ phase jump, and produces a half flux quantum ($\frac{1}{2}\Phi_0=1.03\times10^{-15}Wb$) accompanied by a $\pi$ phase shift for each switching event \cite{2phi_design_Salameh}. These characteristics have fueled additional research into the utilization of $2\phi$ junctions.

\section{Logic Cell Implementation with $2\phi$-JJs}
\label{sec:standard_cell}
\noindent
In the $2\phi$-JJ based design, most cells adhere to the conventional RSFQ cell structures, except inductors, which are replaced by normal JJs (0-JJs.) The $2\phi$-JJ serves as the switching component. Consequently, a logic '1' for this logic family is represented by a half flux quantum pulse, with the voltage$\cdot$time product being half that of a full flux quantum ($\frac{1}{2}\Phi_0=1.03\times10^{-15}Wb$.) This modification eliminates the need for inductors in the cell design, thereby avoiding the disadvantages and inconveniences associated with large inductances in RSFQ logic circuits.

Although the designed inductors are eliminated, parasitic inductors unavoidably exist in every connection. Therefore, an intrinsic $0.5pH$ inductance is assumed for each connection during the design process. According to simulations, the tolerance to parasitic inductance exceeds $100\%$ ($0pH$ to $>1pH$.) Note that all parasitic inductances are omitted in the following schematics to improve the clarity of the figures. 

\begin{figure}[h]
    \includegraphics[width=0.45\textwidth]{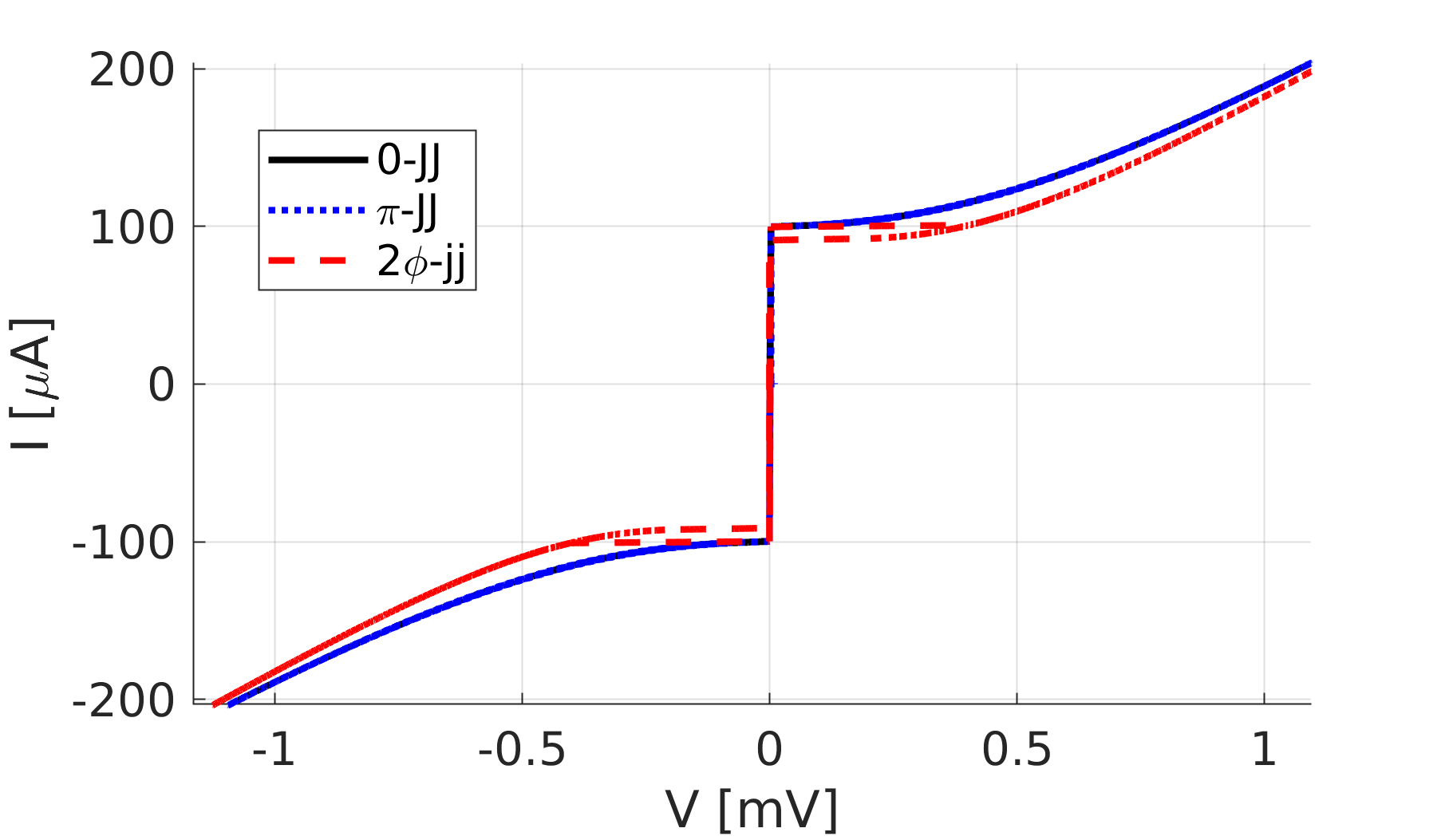}
    \centering
    \caption{Simulation I-V of 0, $\pi$ and $2\phi$ JJs.}
    \centering
    \label{fig:JJtest}
\end{figure}

To confirm the behavior of the JJ models, we simulate the I-V characteristic of the 0, $\pi$, and $2\phi$ JJs \cite{JOINUS} and compare them in Fig.\ref{fig:JJtest}. The difference between 0-JJ and $2\phi$-JJ is in the switching behavior. Each junction switches when the voltage-time integration exceeds the flux quantum requirement. The $2\phi$-JJ switches twice when the 0-JJ switches once while receiving an equivalent amount of flux quantum. The area under the voltage-time graph of a 0-JJ pulse is always $\mathrm{\phi_0}$ whereas $2\phi$-JJs create a pulse with half that value.

In the following sections, we will explore the specifics of each cell, presenting schematic diagrams accompanied by a table listing parameter values and associated cell margins. Additionally, we will provide simulation waveforms to demonstrate the correct functionality of each cell. Preliminary layouts have been generated for these cells to assess potential area savings. However, it's essential to note that, at present, no existing technology offers the $2\phi$-JJ in a fabrication stack-up. Consequently, these layouts cannot be fabricated in any facility known to the authors. For our layout design, we employed the MITLL SFQ5ee process parameters with a simulated $2\phi$-JJ device. To maintain clarity and minimize redundancy, we will only present an example layout of the OR gate in the summary section, providing readers with a preliminary physical view of the cell layouts.

\begin{figure}[!t]
\centering
\noindent\begin{minipage}{\linewidth}
% \begin{figure}[t]
    \includegraphics[width=0.48\textwidth]{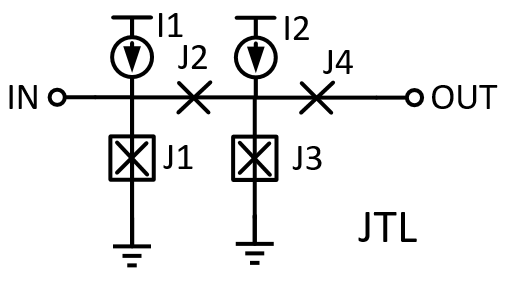}
    \centering
    \captionof{figure}{Schematic of the Josephson Transmission Line (JTL).}
    \centering
    \label{fig:JTL_sch}
% \end{figure}

% \begin{figure}[t]
    \includegraphics[width=0.68\textwidth]{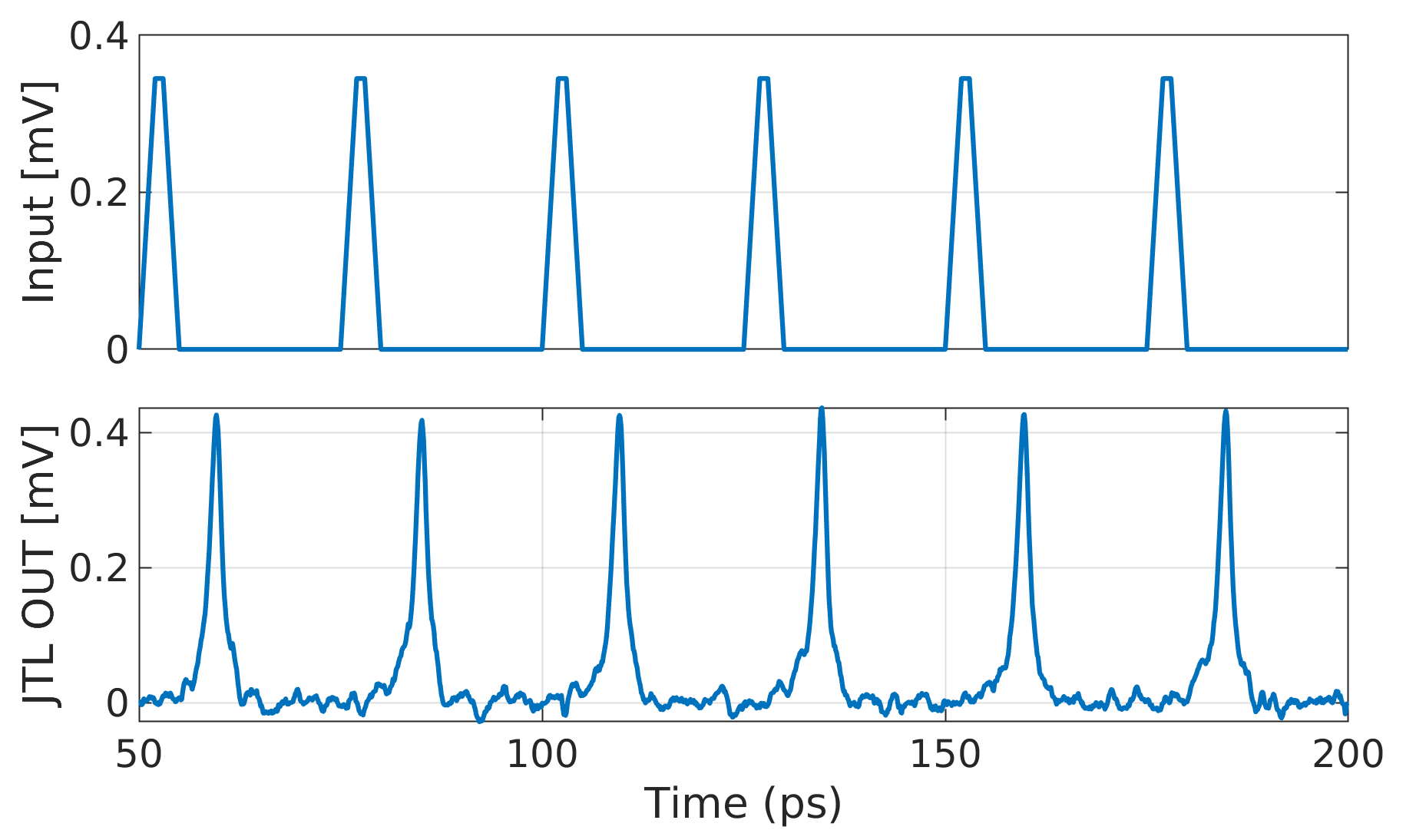}
    \centering
    \captionof{figure}{Simulation waveform of the Josephson Transmission Line (JTL) with added noise.}
    \centering
    \label{fig:JTL_sim}
% \end{figure}

% \begin{table}[!t]
\centering
% \small
\captionof{table}{Parameter values and margins of JTL} 
\label{table:JTL}
\resizebox{0.5\textwidth}{!}{%
\begin{tabular}{|c|c|c|}
\hline
     Components & Values & Margins  \\
     \hline
     J1,J3 & $70\mu A$ & $71\%$\\
     J2,J4 & $80\mu A$ & $97\%$\\
     I1,I2 & $45\mu A$ & $88\%$\\
     Bias  & $1mV$ & $88\%$ \\
\hline
\end{tabular}}
% \end{table}
% \end{minipage}%
% \end{figure}

% \begin{figure}[t]
% \centering
% \noindent\begin{minipage}{\linewidth}
% \begin{figure}[t]
    \includegraphics[width=0.6\textwidth]{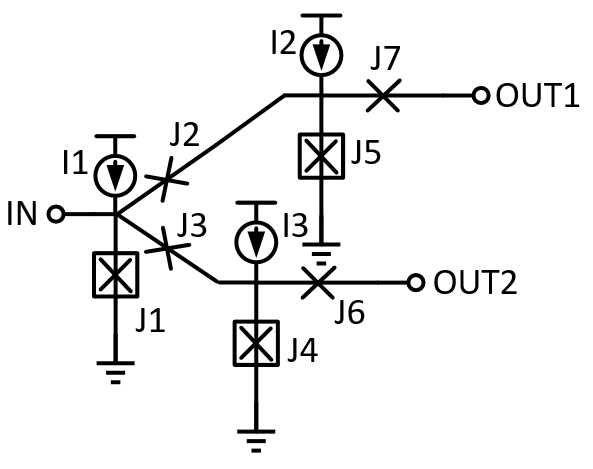}
    \centering
    \captionof{figure}{Schematic of the splitter cell.}
    \centering
    \label{fig:Splitter_sch}
% \end{figure}

% \begin{figure}[t]
    \includegraphics[width=0.76\textwidth]{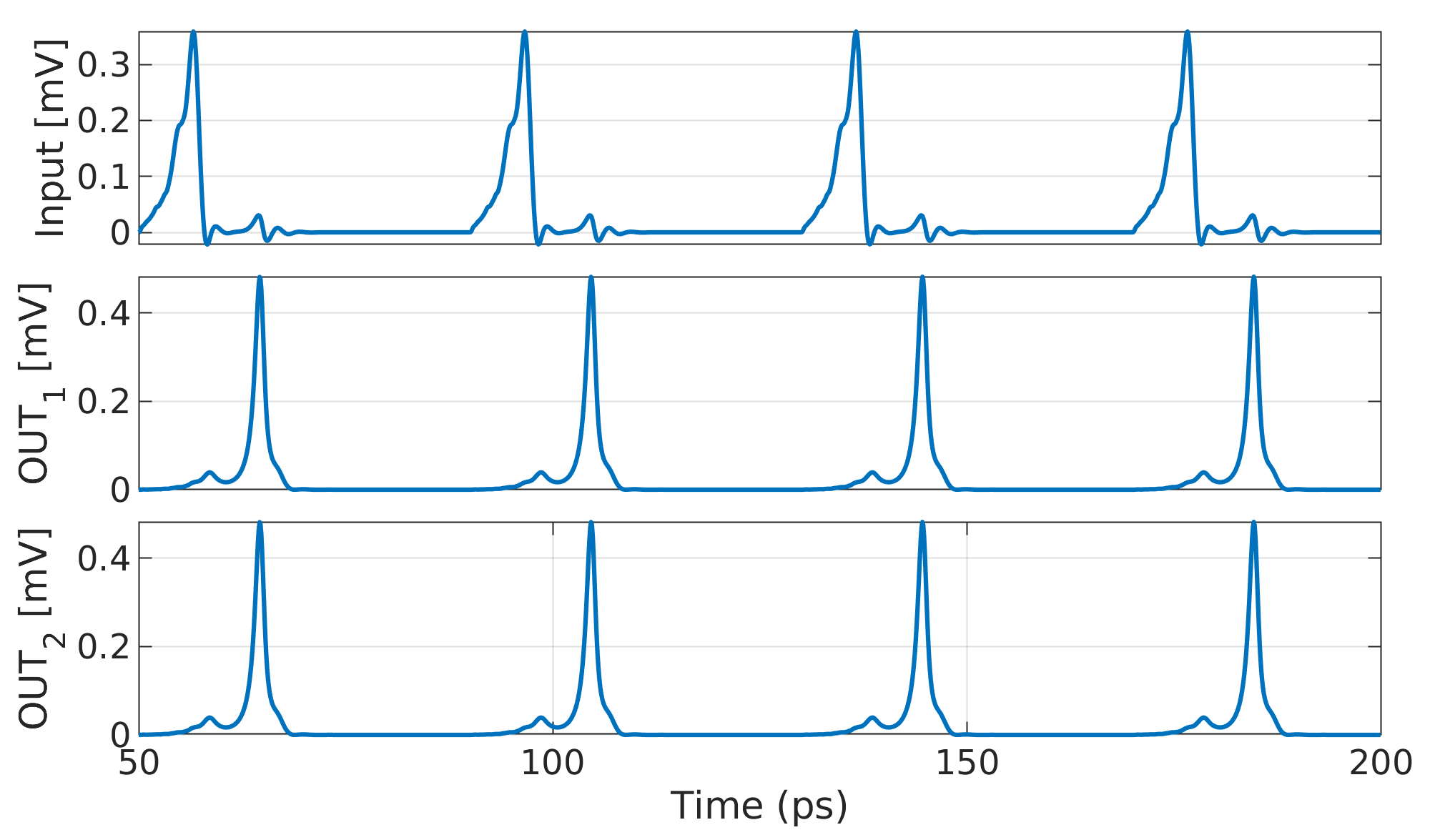}
    \centering
    \captionof{figure}{Simulation waveform of the splitter cell with added noise.}
    \centering
    \label{fig:splitter_sim}
% \end{figure}

% \begin{table}[!h]
\centering
% \small
\captionof{table}{Parameter values and margins of splitter cell} 
\label{table:splitter}
\resizebox{\textwidth}{!}{%
\begin{tabular}{|c|c|c|c|c|c|}
\hline
     Components & Values & Margins & Components & Values & Margins  \\
     \hline
     J1 & $65\mu A$ & $98\%$ & I1 & $70\mu A$ & $100\%$\\
     J2,J3 & $73\mu A$ & $99\%$ & I2,I3 & $45\mu A$ & $70\%$\\
     J4,J5 & $80\mu A$ & $88\%$ & Bias  & $1mV$ & $88\%$\\
     J6,J7 & $80\mu A$ & $90\%$& & & \\
\hline
\end{tabular}}
% \end{table}
\end{minipage}%
\end{figure}
\subsection{Wiring cells}
\subsubsection{Josephson Transmission Line (TP-JTL)}
Fig.\ref{fig:JTL_sch} illustrates the schematic of a JTL block. Within this schematic, devices denoted by square boxes (J1 and J3) represent the $2\phi$-JJs, while those without boxes are 0-JJs (J2 and J4.) Notably, the structure is duplicated, meaning that J1/J3, J2/J4, and I1/I2 are identical. This design ensures the JTL cell's repeatability, allowing for the creation of a JTL chain by connecting the OUT port to a subsequent JTL IN port without introducing unnecessary components.

As depicted, J1 replaces the inductor that is present in the previous RSFQ JTL, forming a J1-J2-J3 loop that establishes a phase equation. This equation ensures that the integration of the phase difference, starting from the positive terminal of J1 through J2, FJ3, and back to J1, amounts to an integer multiple of $2\pi$. When a half-flux-quantum (HFQ) pulse arrives from the IN port, J1 switches and generates another HFQ pulse, which then propagates to the next device. This mechanism facilitates the transmission of the HFQ pulse along the JTL. The simulation waveform is depicted in Fig.\ref{fig:JTL_sim}, and the component values can be found in Table~\ref{table:JTL}. Notably, the critical margin for JTL is 71\% dictated by J1/J3.

\begin{figure}[t]
\centering
\noindent\begin{minipage}{\linewidth}
% \begin{figure}[t]
    \includegraphics[width=0.52\textwidth]{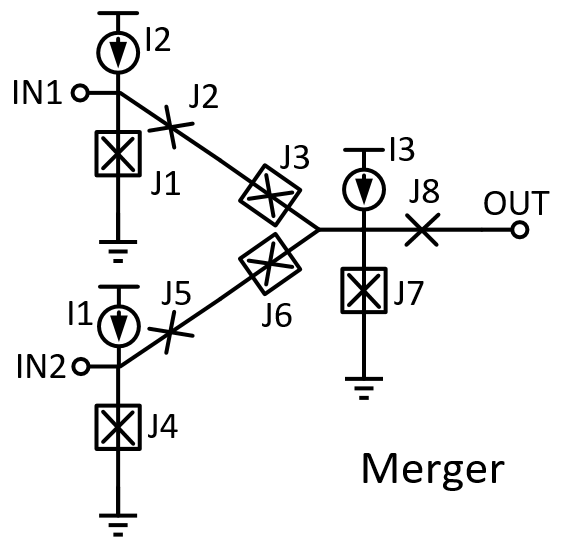}
    \centering
    \captionof{figure}{Schematic of the merger cell.}
    \centering
    \label{fig:Merger_sch}
% \end{figure}

% \begin{figure}[t]
    \includegraphics[width=0.76\textwidth]{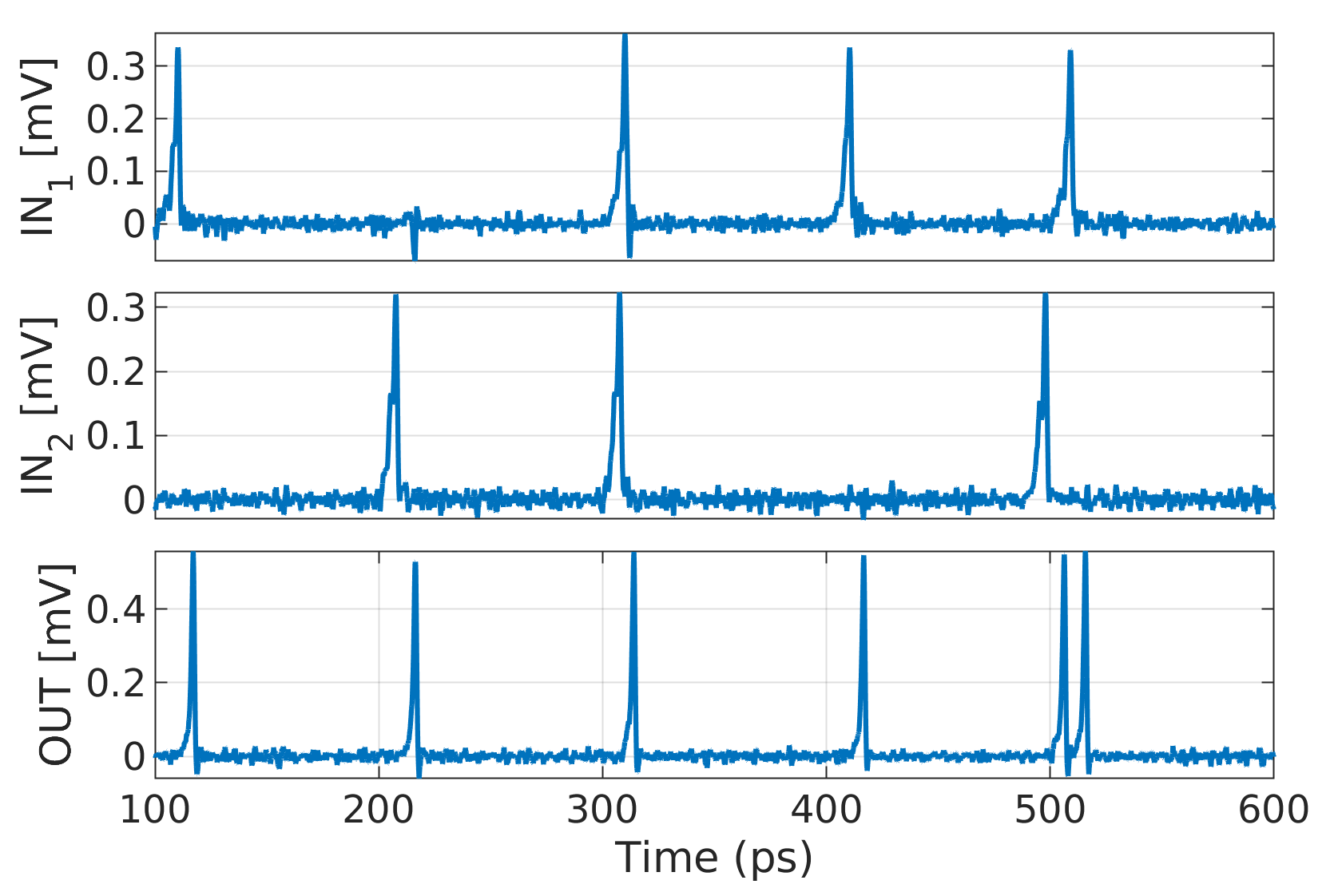}
    \centering
    \captionof{figure}{Simulation waveform of the merger cell with added noise.}
    \centering
    \label{fig:Merger_sim}
% \end{figure}

% \begin{table}[hb]
\centering
\captionof{table}{Parameter values and margins of merger cell} 
\label{table:merger}
\resizebox{\textwidth}{!}{%
\begin{tabular}{|c|c|c|c|c|c|}
\hline
     Components & Values & Margins & Components & Values & Margins \\
     \hline
     J1,J4 & $83\mu A$ & $75\%$ & J8 & $93\mu A$ & $82\%$\\
     J2,J5 & $151\mu A$ & $100\%$ & I1,I2 & $27\mu A$ & $100\%$\\
     J3,J6 & $62\mu A$ & $91\%$ & I3 & $120\mu A$ & $81\%$\\
     J7 & $40\mu A$ & $100\%$ & Bias  & $1mV$ & $72\%$\\
\hline
\end{tabular}}
% \end{table}
\end{minipage}%
\end{figure}

\subsubsection{SPLITTER}
Similar to SFQ cells, HFQ logic cells have a fan-out of one. Therefore, a splitter is employed to duplicate the pulse. Fig.\ref{fig:Splitter_sch} presents the schematic of the splitter cell. In this arrangement, J1 receives the pulse from the IN port, and the looping current is divided into two branches. This division triggers J5 and J4 separately, generating an HFQ pulse at each output port. The simulation waveform is displayed in Fig.\ref{fig:splitter_sim}, and the component values can be found in Table~\ref{table:splitter}. Notably, the critical margin for the splitter is 70\% dictated by the current bias I2/I3.

\subsubsection{MERGER}
Fig. \ref{fig:Merger_sch} illustrates the schematic of the merger cell, often referred to as a confluence buffer. When it receives an HFQ pulse from either input port (e.g., IN1), the pulse triggers the corresponding junction (J1). This action increases the current in the respective branch (J1-J2-J3-J7), leading to the switching of J7, resulting in an output pulse. Concurrently, a buffering junction on the other branch (in this case, J6 for input from IN1) is triggered to counteract the backward flux flow towards the other input port (IN2). Specifically, J3 and J6 serve as barriers to prevent the reverse propagation of pulses. In the event that two input pulses arrive simultaneously or within a small time window (several picoseconds), only one output pulse is generated at the OUT port. The simulation waveform is depicted in Fig. \ref{fig:Merger_sim}, illustrating both the input and output waveforms and the phases of J3 and J6 to demonstrate how they inhibit backpropagation. Detailed component values can be found in Table \ref{table:merger}. It's worth noting that the critical margin for the merger is 72\%, determined by the overall bias voltage.

\begin{figure}[t]
\centering
\noindent\begin{minipage}{\linewidth}
% \begin{figure}[t]
    \includegraphics[width=0.44\textwidth]{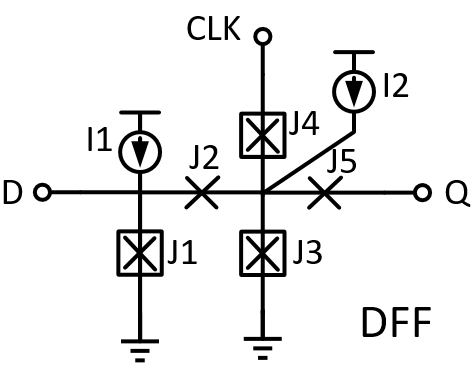}
    \centering
    \captionof{figure}{Schematic of the D Flip-flop.}
    \centering
    \label{fig:DFF_sch}
% \end{figure}

% \begin{figure}[t]
    \includegraphics[width=0.76\textwidth]{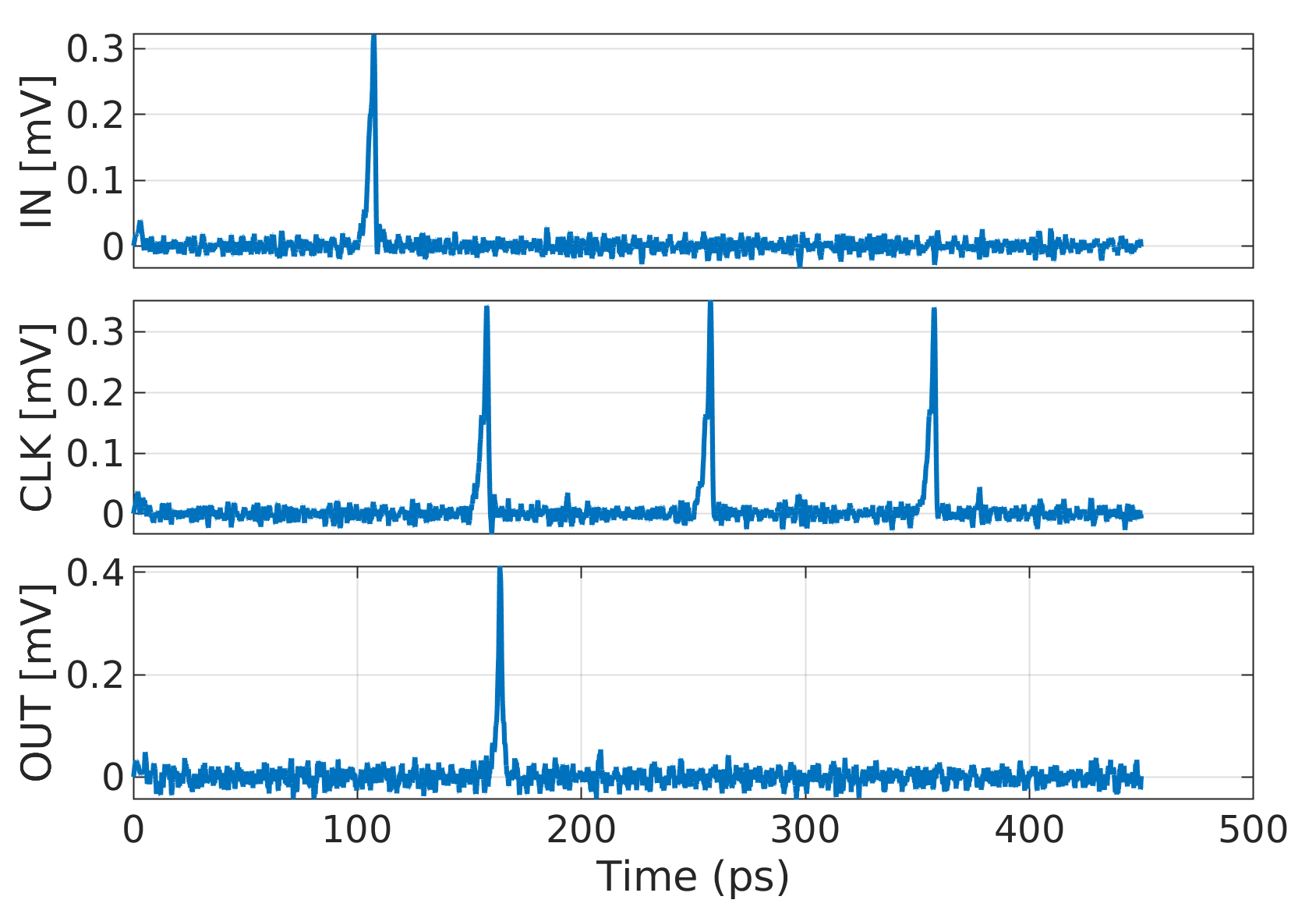}
    \centering
    \captionof{figure}{Simulation waveform of the D Flip-flop with added noise.}
    \centering
    \label{fig:DFF_sim}
% \end{figure}

% \begin{table}[b]
\centering
\captionof{table}{Parameter values and margins of DFF} 
\label{table:DFF}
\resizebox{\textwidth}{!}{%
\begin{tabular}{|c|c|c|c|c|c|}
\hline
     Components & Values & Margins & Components & Values & Margins \\
     \hline
     J1 & $63\mu A$ & $68\%$ & J5 & $117\mu A$&$97\%$\\
     J2 & $65\mu A$ & $90\%$ & I1 & $31\mu A$&$100\%$\\
     J3 & $86\mu A$ & $75\%$ & I2 & $32\mu A$&$100\%$\\
     J4 & $74\mu A$&$85\%$ & Bias  & $1mV$ & $88\%$\\
\hline
\end{tabular}}
% \end{table}
\end{minipage}%
\end{figure}

\subsection{Memory cells}
\subsubsection{D FLIP-FLOP}
\noindent
Fig.\ref{fig:DFF_sch} illustrates the schematic of the DFF (Data Flip-Flop.) When an HFQ pulse arrives from the input port D, it is stored within the J1-J2-J3 loop as a clockwise looping current. This action increases the bias current of J3. Consequently, when an HFQ pulse arrives from the CLK port, it triggers J3, generating an HFQ pulse at the output port Q. In cases where no HFQ is stored, the incoming pulse from CLK triggers the J4 junction, resulting in no output at Q. The simulation waveform is displayed in Fig.\ref{fig:DFF_sim}, and the component values are detailed in Table.\ref{table:DFF}. Notably, the critical margin for the DFF is 68\% dictated by J1.

\subsubsection{NDRO}
\begin{figure}[t]
\centering
\noindent\begin{minipage}{\linewidth}
% \begin{figure}[h]
    \includegraphics[width=0.52\textwidth]{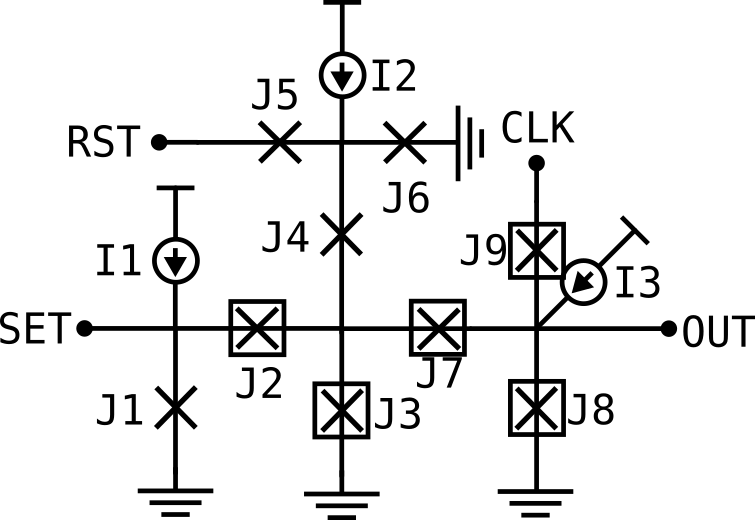}
    \centering
    \captionof{figure}{Schematic of the 2$\phi$-based NDRO memory cell.}
    \centering
    \label{fig:NDRO_sch}
% \end{figure}

% \begin{figure}[h]
    \includegraphics[width=0.8\textwidth]{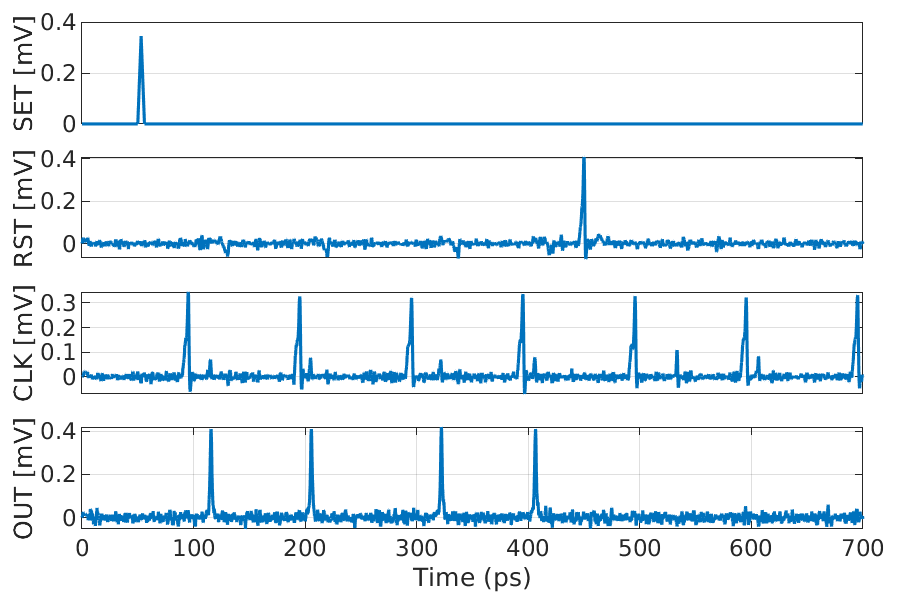}
    \centering
    \captionof{figure}{Simulation waveform of the NDRO cell with added noise.}
    \centering
    \label{fig:NDRO_sim}
% \end{figure}

% \begin{table}[h]
\centering
\captionof{table}{Parameter values and margins of NDRO cell} 
\label{table:NDRO}
\resizebox{\textwidth}{!}{%
\begin{tabular}{|c|c|c|c|c|c|}
\hline
     Components & Values & Margins & Components & Values & Margins \\
     \hline
     J1 & $75\mu A$ & $100\%$ & J7 & $58\mu A$ & $100\%$\\
     J2, & $63\mu A$ & $58\%$ & J8 & $48\mu A$ & $70\%$\\
     J3, & $44\mu A$ & $94\%$ & J9 &  $107\mu A$ & $76\%$\\
     J4 & $139\mu A$ & $94\%$ & I1  & $236\mu A$ & $100\%$\\
     J5 & $88\mu A$ & $57\%$ & I2  & $123\mu A$ & $70\%$\\
     J6 & $56\mu A$ & $97\%$ & I3  & $128\mu A$ & $75\%$\\
\hline
\end{tabular}}
% \end{table}
\end{minipage}%
\end{figure}

\noindent
Non-destructive memory cells are widely used as a memory unit in SFQ circuits. The DFF cell is a traditional Destructive Read-Out (DRO) unit that stores a single bit of information. Its core architecture is a simple storage loop that stores flux from input and releases it to output by a clock signal. In contrast to the DRO, NDRO does not release the pulse by the clock signal and requires a reset pin to clear the stored value within its storage loop. Fig.\ref{fig:NDRO_sch} shows the structure of the NDRO cell. The NDRO design values are shown in Table.\ref{table:NDRO}. The simulation waveform can be observed in Fig.\ref{fig:NDRO_sim}.
\subsection{Logic cells}
\begin{figure}[!ht]
\centering
\noindent\begin{minipage}{\linewidth}
% \begin{figure}[h]
    \includegraphics[width=0.52\textwidth]{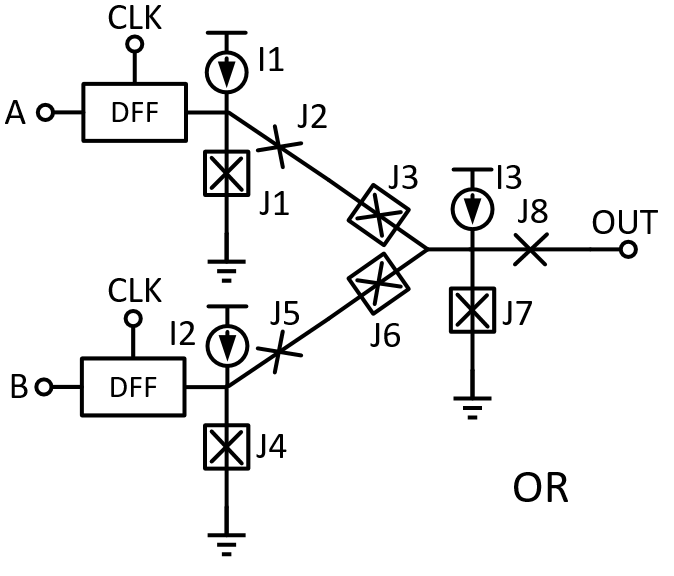}
    \centering
    \captionof{figure}{Schematic of the OR gate.}
    \centering
    \label{fig:OR_sch}
% \end{figure}

% \begin{figure}[h]
    \includegraphics[width=0.78\textwidth]{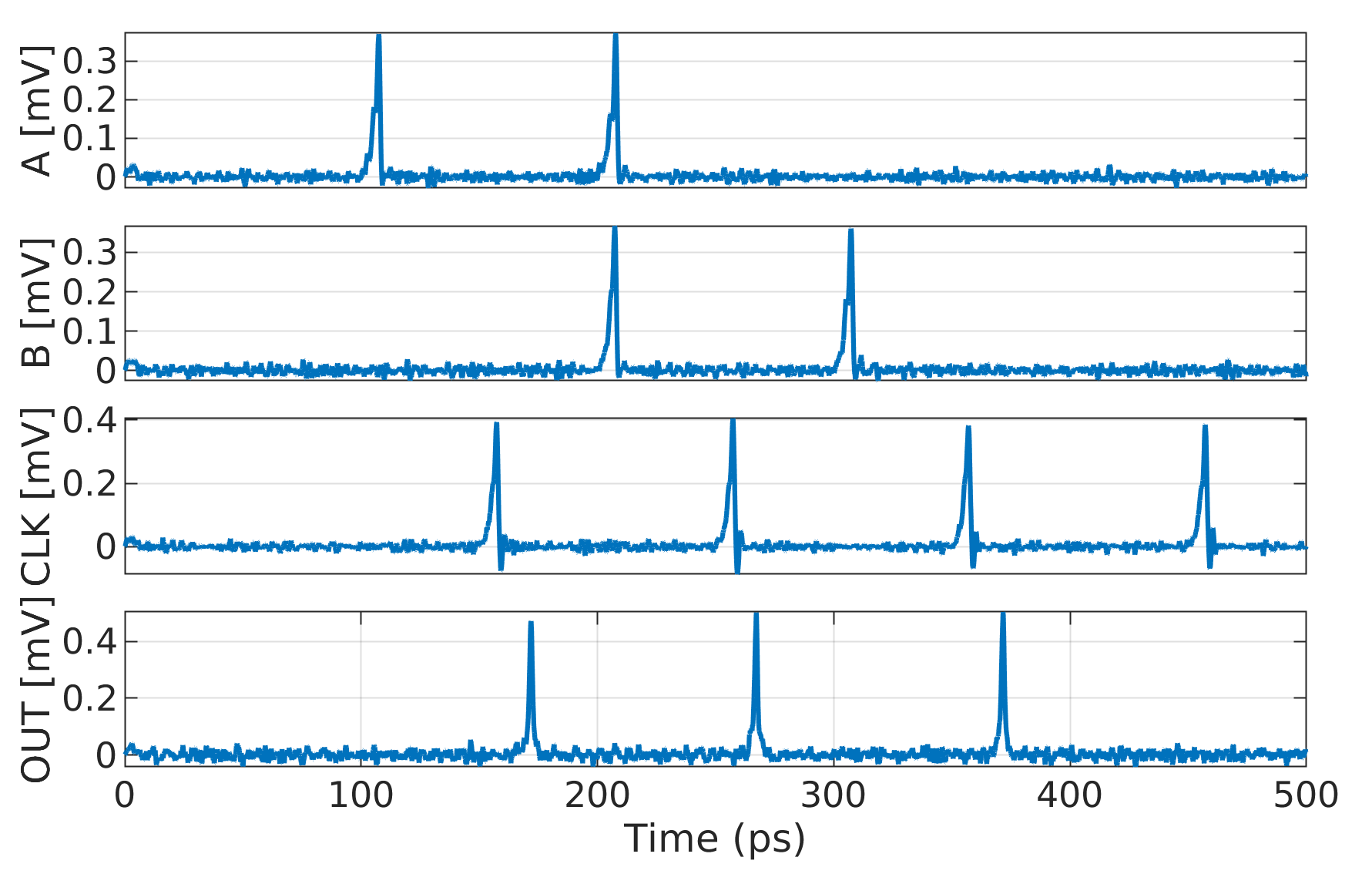}
    \centering
    \captionof{figure}{Simulation waveform of the OR gate.}
    \centering
    \label{fig:OR_sim}
% \end{figure}

% \begin{table}[h]
\centering
\captionof{table}{Parameter values and margins of OR gate} 
\label{table:OR}
\resizebox{\textwidth}{!}{%
\begin{tabular}{|c|c|c|c|c|c|}
\hline
     Components & Values & Margins & Components & Values & Margins \\
     \hline
     J1,J4 & $95\mu A$ & $40\%$ & J8 & $74\mu A$ & $69\%$\\
     J2,J5 & $100\mu A$ & $67\%$ & I1,I2 & $25\mu A$ & $93\%$\\
     J3,J6 & $83\mu A$ & $37\%$ & I3 &  $83\mu A$ & $99\%$\\
     J7 & $68\mu A$ & $73\%$ & Bias  & $1mV$ & $44\%$\\
\hline
\end{tabular}}
% \end{table}
% \end{minipage}%
% \end{figure}

% \begin{figure}[t]
% \centering
% \noindent\begin{minipage}{\linewidth}
% \begin{figure}[h]
    \includegraphics[width=0.52\textwidth]{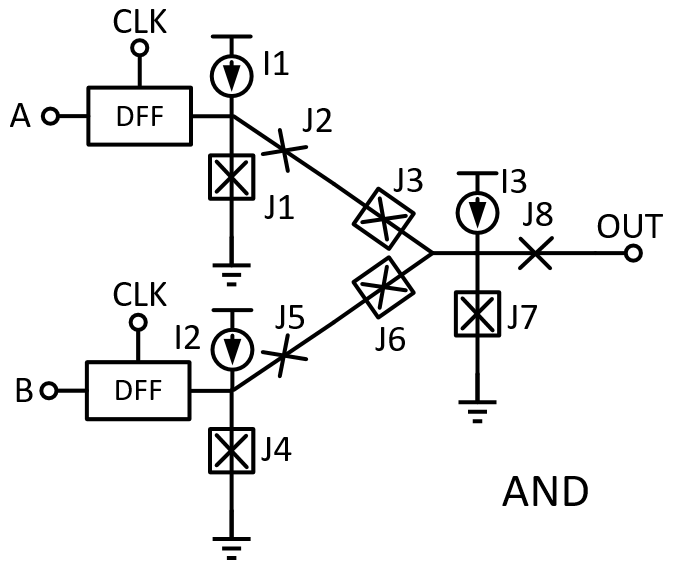}
    \centering
    \captionof{figure}{Schematic of the AND gate.}
    \centering
    \label{fig:AND_sch}
% \end{figure}
%
% \begin{figure}[h]
    \includegraphics[width=0.78\textwidth]{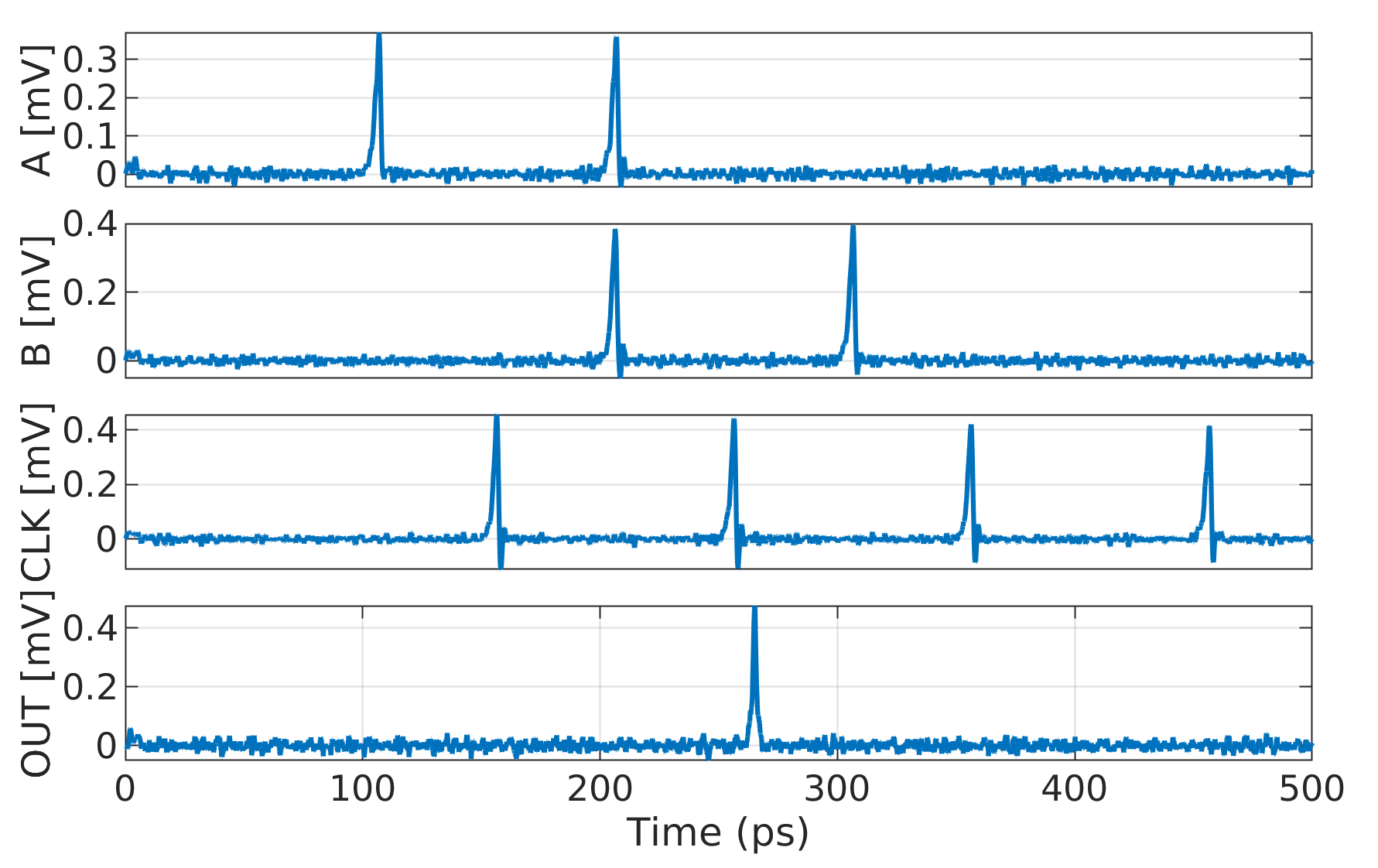}
    \centering
    \captionof{figure}{Simulation waveform of the AND gate.}
    \centering
    \label{fig:AND_sim}
% \end{figure}
%
% \begin{table}[h]
\centering
\captionof{table}{Parameter values and margins of AND gate} 
\label{table:AND}
\resizebox{\textwidth}{!}{%
\begin{tabular}{|c|c|c|c|c|c|}
\hline
     Components & Values & Margins & Components & Values & Margins \\
     \hline
     J1,J4 & $70\mu A$ & $97\%$ & J8 & $74\mu A$ & $85\%$\\
     J2,J5 & $140\mu A$ & $99\%$ & I1,I2 & $31\mu A$ & $100\%$\\
     J3,J6 & $72\mu A$ & $94\%$ & I3 &  $32\mu A$ & $100\%$\\
     J7 & $99\mu A$ & $98\%$ & Bias  & $1mV$ & $92\%$\\
\hline
\end{tabular}}
% \end{table}
\end{minipage}%
\end{figure}

\subsubsection{TP-OR}
\noindent
To grasp the concept of the OR gate's operation, consider it as two DFFs driven by the same clock and followed by a merger, as depicted in Fig.\ref{fig:OR_sch}. The DFF cell has already been discussed, rendering it unnecessary to reiterate its details. The subsequent merging section possesses the same structure as the previously introduced merger. However, there are slight differences in component values as the tool automatically optimizes them. These optimized values are displayed in Table.\ref{table:OR}, with the critical margin being 40\% dictated by J1/J4. The simulation waveform can be observed in Fig.\ref{fig:OR_sim}.

\subsubsection{TP-AND}
\noindent
As depicted in Fig.\ref{fig:AND_sch}, the AND gate adopts an identical structure to the OR gate. However, it manipulates the key components, specifically the merging part, to ensure that the output junction J7 necessitates at least two HFQ pulses to generate an output pulse.
The simulation waveform is illustrated in Fig.\ref{fig:AND_sim}, and the component values are provided in Table.\ref{table:AND}. Notably, the critical margin for the merger is 85\%, dictated by J8.

\subsubsection{TP-XOR}
\begin{figure}[t]
\centering
\noindent\begin{minipage}{\linewidth}
% \begin{figure}[h]
    \includegraphics[width=0.6\textwidth]{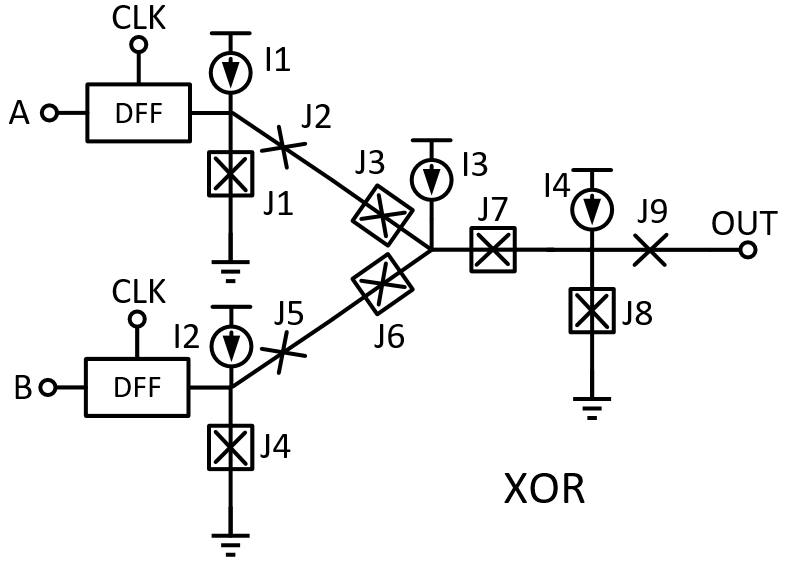}
    \centering
    \captionof{figure}{Schematic of the XOR gate.}
    \centering
    \label{fig:XOR_sch}
% \end{figure}

% \begin{figure}[h]
    \includegraphics[width=0.8\textwidth]{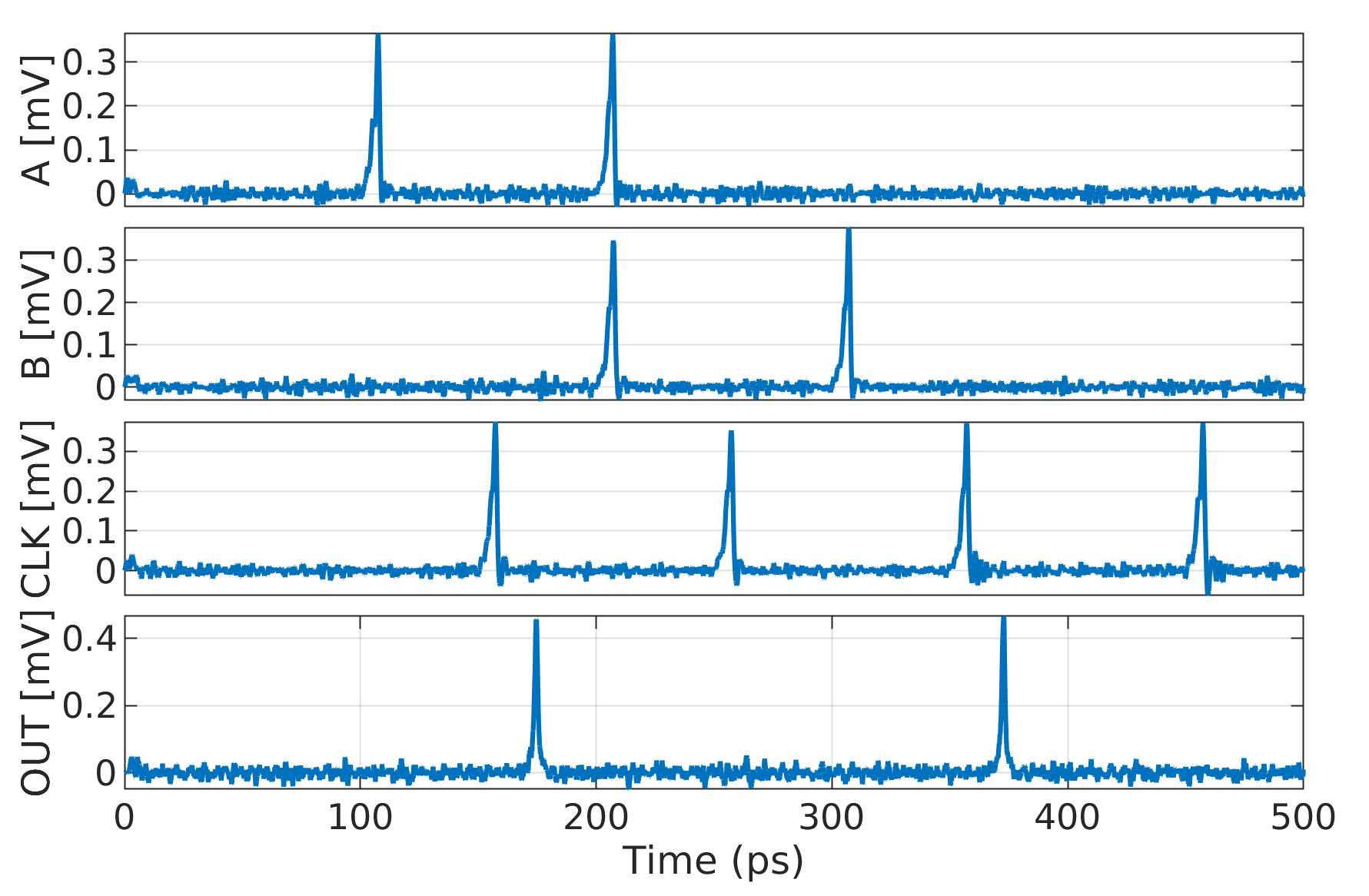}
    \centering
    \captionof{figure}{Simulation waveform of the XOR gate.}
    \centering
    \label{fig:XOR_sim}
% \end{figure}

% \begin{table}[h]
\centering
\captionof{table}{Parameter values and margins of XOR gate} 
\label{table:XOR}
\resizebox{\textwidth}{!}{%
\begin{tabular}{|c|c|c|c|c|c|}
\hline
     Components & Values & Margins & Components & Values & Margins \\
     \hline
     J1,J4 & $87\mu A$ & $46\%$ &J9 & $80\mu A$ & $89\%$\\
     J2,J5 & $95\mu A$ & $75\%$ &I1,I2 & $31\mu A$ & $79\%$\\
     J3,J6 & $72\mu A$ & $40\%$&I3 &  $63\mu A$ & $88\%$\\
     J7 & $70\mu A$ & $20\%$ &I4 &  $39\mu A$ & $39\%$\\
     J8 & $72\mu A$ & $22\%$&Bias  & $1mV$ & $26\%$\\
\hline
\end{tabular}}
% \end{table}
\end{minipage}%
\end{figure}

\noindent
The schematic of the XOR gate is illustrated in Fig.\ref{fig:XOR_sch}. In contrast to the AND or OR gate, there is an additional junction (J7) after the merging point of the two input branches. With J7 in place, when pulses arrive from both branches (corresponding to the case: A=1 and B=1), J7 switches, while J8 remains inactive, resulting in no output pulse. Conversely, in situations where only one pulse is received from either of the branches (corresponding to the cases: A=1, B=0, or A=0, B=1), J8 is triggered, producing the output pulse. The simulation waveform is visualized in Fig.\ref{fig:XOR_sim}, and the component values can be found in Table.\ref{table:XOR}. Notably, the critical margin for the merger is 20\%, dictated by J7.

\subsubsection{TP-INV}
\begin{figure}[t]
\centering
\noindent\begin{minipage}{\linewidth}
% \begin{figure}[h]
    \includegraphics[width=0.8\textwidth]{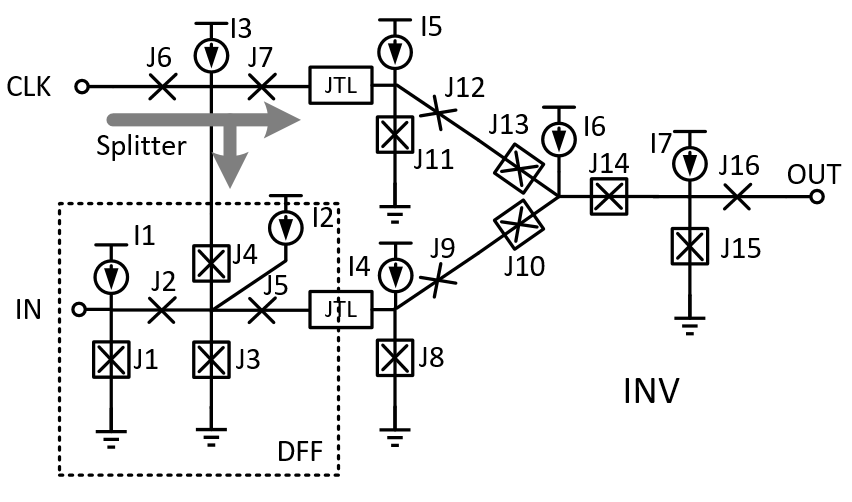}
    \centering
    \captionof{figure}{Schematic of the inverter.}
    \centering
    \label{fig:INV_sch}
% \end{figure}

% \begin{figure}[h]
    \includegraphics[width=0.8\textwidth]{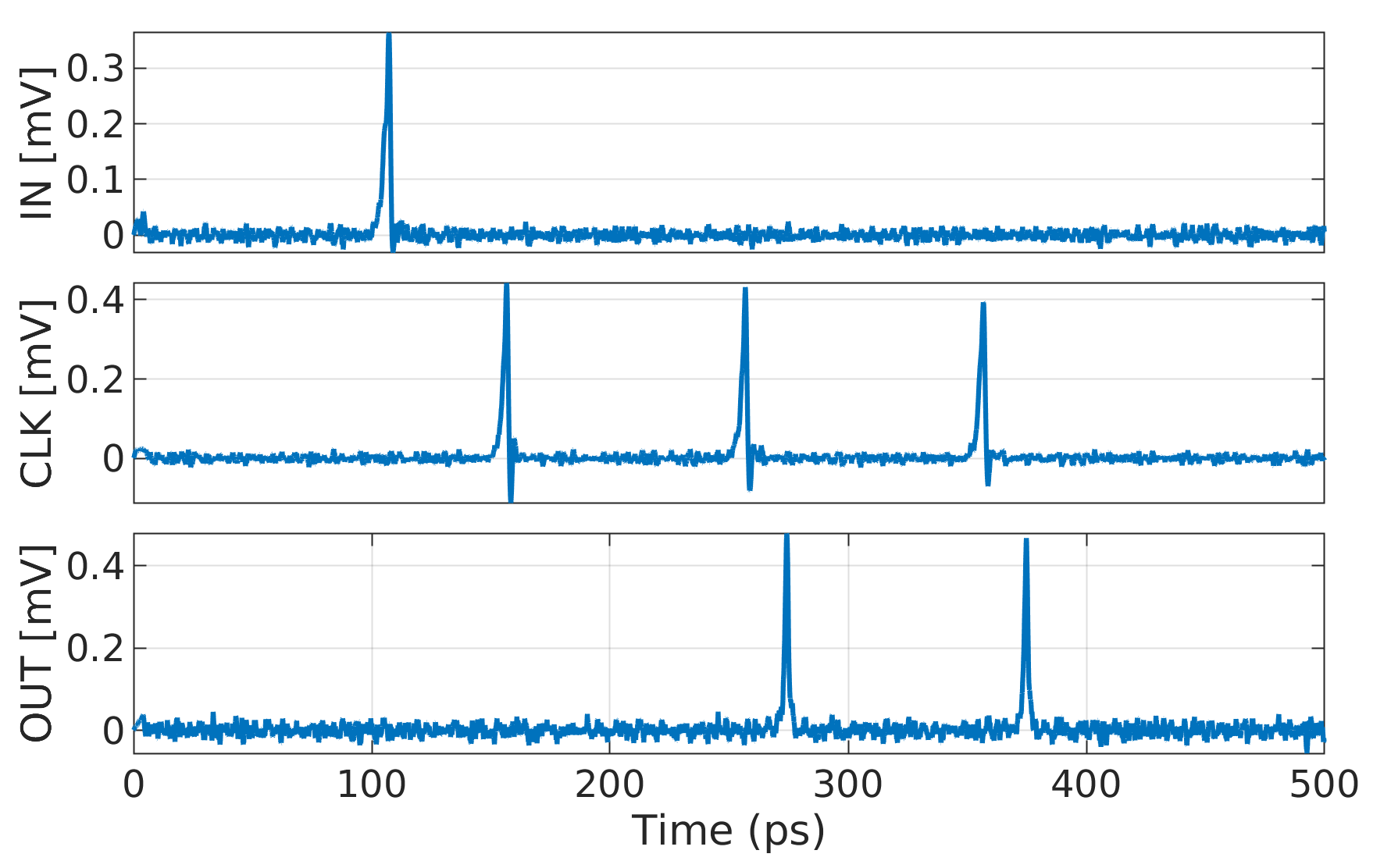}
    \centering
    \captionof{figure}{Simulation waveform of the inverter.}
    \centering
    \label{fig:INV_sim}
% \end{figure}
%
% \begin{table}[h]
\centering
\captionof{table}{Parameter values and margins of the inverter} 
\label{table:INV}
\resizebox{\textwidth}{!}{%
\begin{tabular}{|c|c|c|c|c|c|}
\hline
     Components&Values&Margins&Components&Values&Margins\\
     \hline
     J1 & $85\mu A$ & $43\%$&
     J2 & $77\mu A$ & $69\%$\\
     J3 & $95\mu A$ & $52\%$&
     J4 & $70\mu A$ & $48\%$\\
     J5 & $82\mu A$ & $62\%$&
     J6 & $80\mu A$ & $81\%$\\
     J7 & $80\mu A$ & $61\%$&
     J8,J11 & $96\mu A$ & $76\%$\\
     J9,J12 & $101\mu A$ & $85\%$&
     J10,J13 & $78\mu A$ & $64\%$\\
     J14 & $80\mu A$ & $16\%$&
     J15 & $82\mu A$ & $16\%$\\
     J16 & $90\mu A$ & $67\%$&
     I1 & $43\mu A$ & $63\%$\\
     I2 & $40\mu A$ & $79\%$&
     I3 &  $57\mu A$ & $65\%$\\
     I4,I5 &  $37\mu A$ & $79\%$&
     I6 &  $70\mu A$ & $73\%$\\
     I7 &  $40\mu A$ & $36\%$&
     Bias  & $1mV$ & $22\%$ \\
\hline
\end{tabular}}
% \end{table}
\end{minipage}%
\end{figure}

\noindent
Fig.\ref{fig:INV_sch} displays the schematic of an inverter. The upper section of the cell has a splitter-like structure, but one branch is merged with the clock branch of a DFF-like structure at the bottom. Consequently, when the clock signal arrives, the first part of the inverter generates '11' or '10' based on whether an input pulse was stored. Subsequently, the following XOR-like structure completes the inversion operation. The simulation waveform is depicted in Fig.\ref{fig:INV_sim}, and the component values are provided in Table.\ref{table:INV}. Notably, the critical margin for the merger is 16\%, dictated by J14 and J15.

\subsection{Interface cells}
\subsubsection{PTL Driver AND Receiver}
\begin{figure}[!t]
\centering
\noindent\begin{minipage}{\linewidth}
% \begin{figure}[h]
    \includegraphics[width=0.76\textwidth]{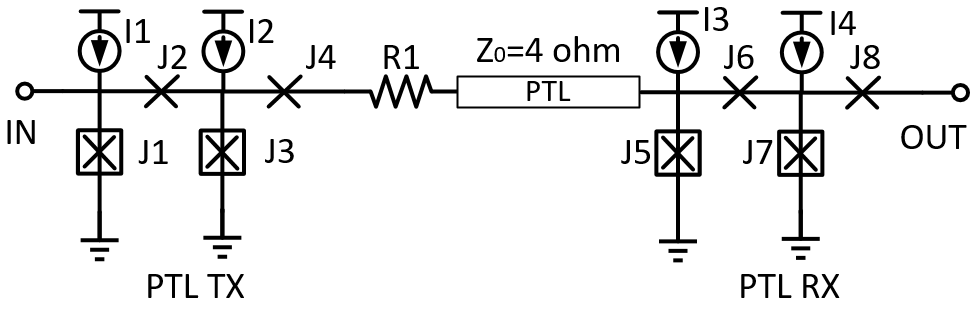}
    \centering
    \captionof{figure}{Schematic of the PTL driver and receiver.}
    \centering
    \label{fig:PTLTRX_sch}
% \end{figure}

% \begin{figure}[h]
    \includegraphics[width=0.72\textwidth]{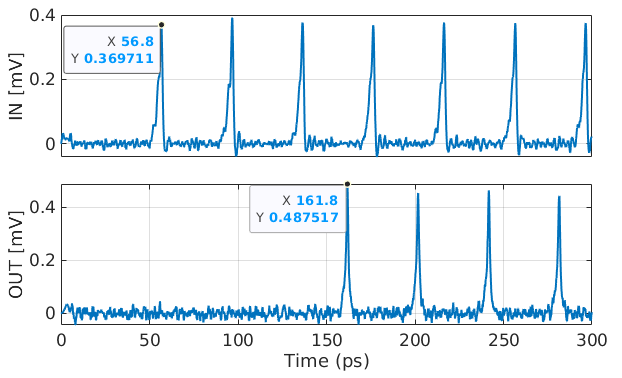}
    \centering
    \captionof{figure}{Simulation waveform of the PTL driver and receiver.}
    \centering
    \label{fig:PTL_sim}
% \end{figure}

% \begin{table}[h]
\centering
\captionof{table}{Parameter values and margins of PTL driver and receive} 
\label{table:PTL}
\resizebox{\textwidth}{!}{%
\begin{tabular}{|c|c|c|c|c|c|}
\hline
     Components & Values & Margins & Components & Values & Margins \\
     \hline
     J1 & $80\mu A$ & $81\%$&J5 & $75\mu A$ & $88\%$\\
     J2 & $80\mu A$ & $84\%$&J6 & $61\mu A$ & $87\%$\\
     J3 & $70\mu A$ & $100\%$&J7 & $80\mu A$ & $81\%$\\
     J4 & $83\mu A$ & $100\%$&J8 & $80\mu A$ & $84\%$\\
     R1 & $0.5 ohms$ & $100\%$&I3 &  $62\mu A$ & $81\%$\\
     I1 & $45\mu A$ & $88\%$&I4 &  $45\mu A$ & $88\%$\\
     I1 & $49\mu A$ & $94\%$&RX bias  & $1mV$ & $88\%$ \\
     TX bias & $1mV$ & $100\%$&  &  & \\
\hline
\end{tabular}}
% \end{table}
% \end{minipage}%
% \end{figure}

% \begin{figure}[t]
% \centering
% \noindent\begin{minipage}{\linewidth}
% \begin{figure}[h]
    \includegraphics[width=0.48\textwidth]{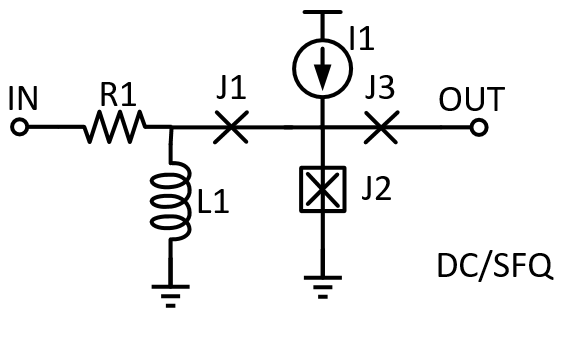}
    \centering
    \captionof{figure}{Schematic of the DC/SFQ converter.}
    \centering
    \label{fig:DCSFQ_sch}
% \end{figure}

% \begin{figure}[h]
    \includegraphics[width=0.66\textwidth]{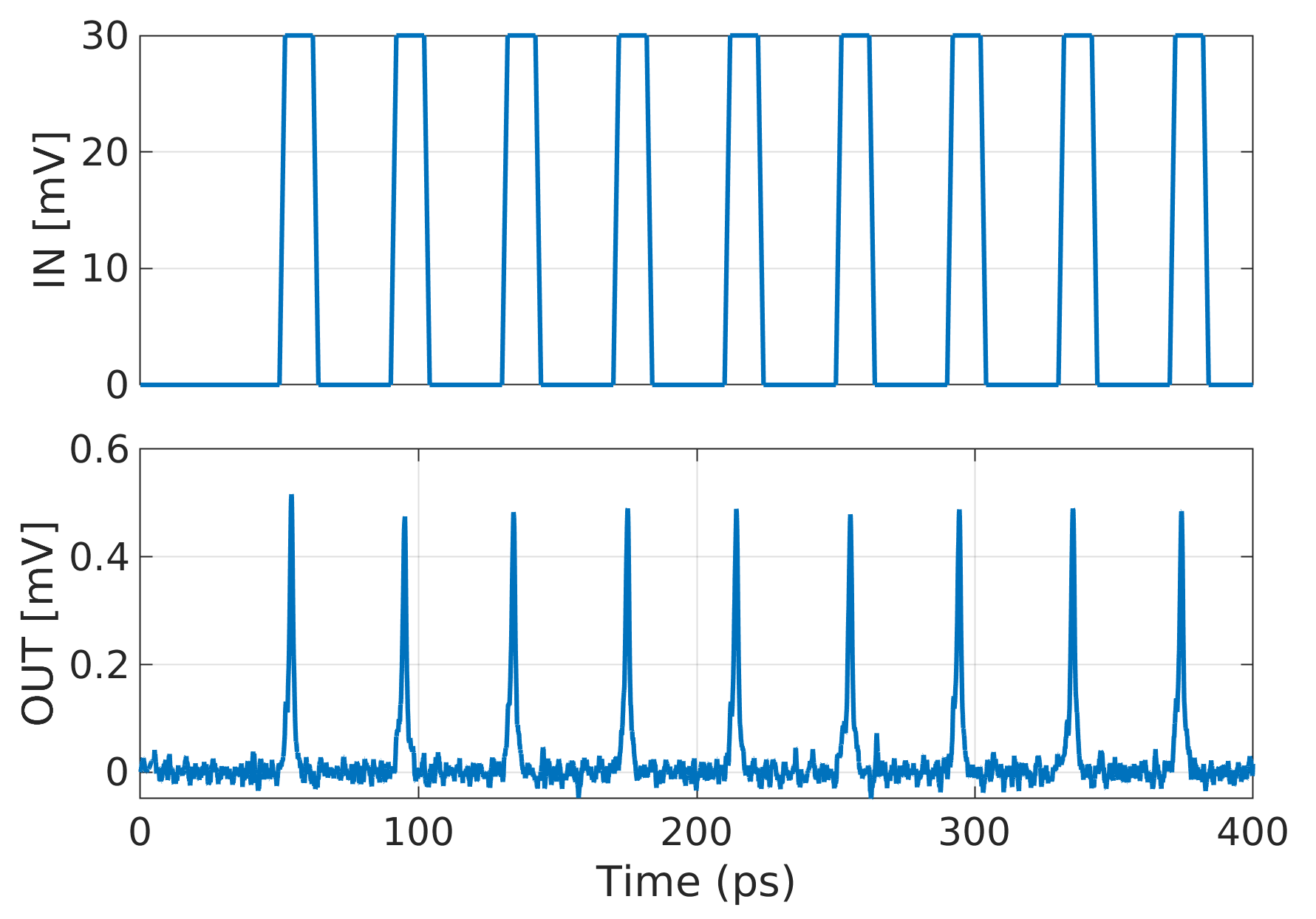}
    \centering
    \captionof{figure}{Simulation waveform of the DC/SFQ converter.}
    \centering
    \label{fig:DCSFQ_sim}
% \end{figure}

% \begin{table}[h]
\centering
\captionof{table}{Parameter values and margins of DC/SFQ converter} 
\label{table:DCSFQ}
\resizebox{\textwidth}{!}{%
\begin{tabular}{|c|c|c|c|c|c|}
\hline
     Components & Values & Margins & Components & Values & Margins \\
     \hline
     R1& $50ohms$ & $100\%$&I1 & $48\mu A$ & $100\%$\\
     J1& $83\mu A$ & $98\%$&L1 &  $6.7pH$ & $100\%$\\
     J2& $71\mu A$ & $100\%$&Bias  & $1mV$ & $100\%$\\
     J3& $100\mu A$ & $100\%$&&&\\
\hline
\end{tabular}}
% \end{table}
\end{minipage}%
\end{figure}

\noindent
Fig.\ref{fig:PTLTRX_sch} presents the schematic for both the PTL transmitter/driver (TX) and receiver (RX). These structures have JTL-like designs, with the PTL driver featuring a serial resistor at the output port for impedance matching. In this configuration, the characteristic impedance is set to four ohms, although designers can select practical values for their technology and adjust circuit parameters accordingly. The simulation waveform is displayed in Fig.\ref{fig:PTL_sim}, and the component values can be found in Table.\ref{table:XOR}. It's worth noting that the critical margin for the PTL driver is 81\%, dictated by J1, and the critical margin for the PTL receiver is 81\%, dictated by J7.

\subsubsection{DC/SFQ CONVERTER}
\noindent
Fig.\ref{fig:DCSFQ_sch} is the DC/SFQ converter schematic. R1 is a serial input resistor which converts the input voltage to current. The resistor can be implemented on-chip (this design) or off-chip. A large inductor L1 follows the input resistor. At the rising edge of the input, L1 reveals high impedance, and most of the current will flow through J2, which triggers an HFQ pulse at the output port. When the input voltage becomes steady, L1 is equivalent to a short connection, and the input current will flow through L1 to the ground, leaving J2 untouched. The simulation waveform is shown in Fig.\ref{fig:DCSFQ_sim}, and the component values are listed in Table.\ref{table:DCSFQ}. The critical margin of the PTL driver is $98\%$ on J1.

% \begin{figure}[t]
% \centering
% \noindent\begin{minipage}{\linewidth}
% % \begin{figure}[h]
%     \includegraphics[width=0.48\textwidth]{figures/DCSFQ_sch.PNG}
%     \centering
%     \captionof{figure}{Schematic of the DC/SFQ converter.}
%     \centering
%     \label{fig:DCSFQ_sch}
% % \end{figure}

% % \begin{figure}[h]
%     \includegraphics[width=0.66\textwidth]{figures/DCSFQ_sim.png}
%     \centering
%     \captionof{figure}{Simulation waveform of the DC/SFQ converter.}
%     \centering
%     \label{fig:DCSFQ_sim}
% % \end{figure}

% % \begin{table}[h]
% \centering
% \captionof{table}{Parameter values and margins of DC/SFQ converter} 
% \label{table:DCSFQ}
% \resizebox{\textwidth}{!}{%
% \begin{tabular}{|c|c|c|c|c|c|}
% \hline
%      Components & Values & Margins & Components & Values & Margins \\
%      \hline
%      R1& $50ohms$ & $100\%$&I1 & $48\mu A$ & $100\%$\\
%      J1& $83\mu A$ & $98\%$&L1 &  $6.7pH$ & $100\%$\\
%      J2& $71\mu A$ & $100\%$&Bias  & $1mV$ & $100\%$\\
%      J3& $100\mu A$ & $100\%$&&&\\
% \hline
% \end{tabular}}
% % \end{table}
% \end{minipage}%
% \end{figure}

\subsubsection{SFQ/DC CONVERTER}
\begin{figure}[t]
\centering
\noindent\begin{minipage}{\linewidth}
% \begin{figure}[h]
    \includegraphics[width=0.68\textwidth]{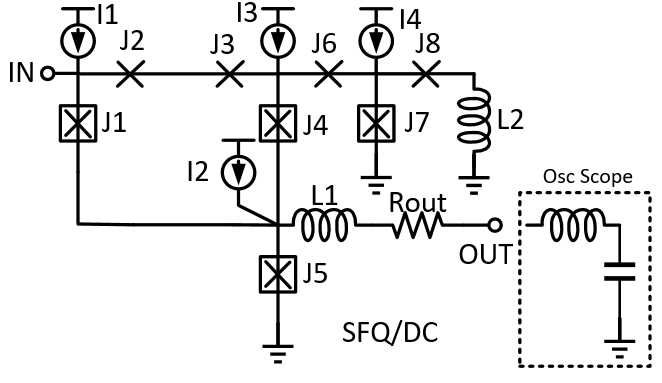}
    \centering
    \captionof{figure}{Schematic of the SFQ/DC converter.}
    \centering
    \label{fig:SFQDC_sch}
% \end{figure}

% \begin{figure}[h]
    \includegraphics[width=0.76\textwidth]{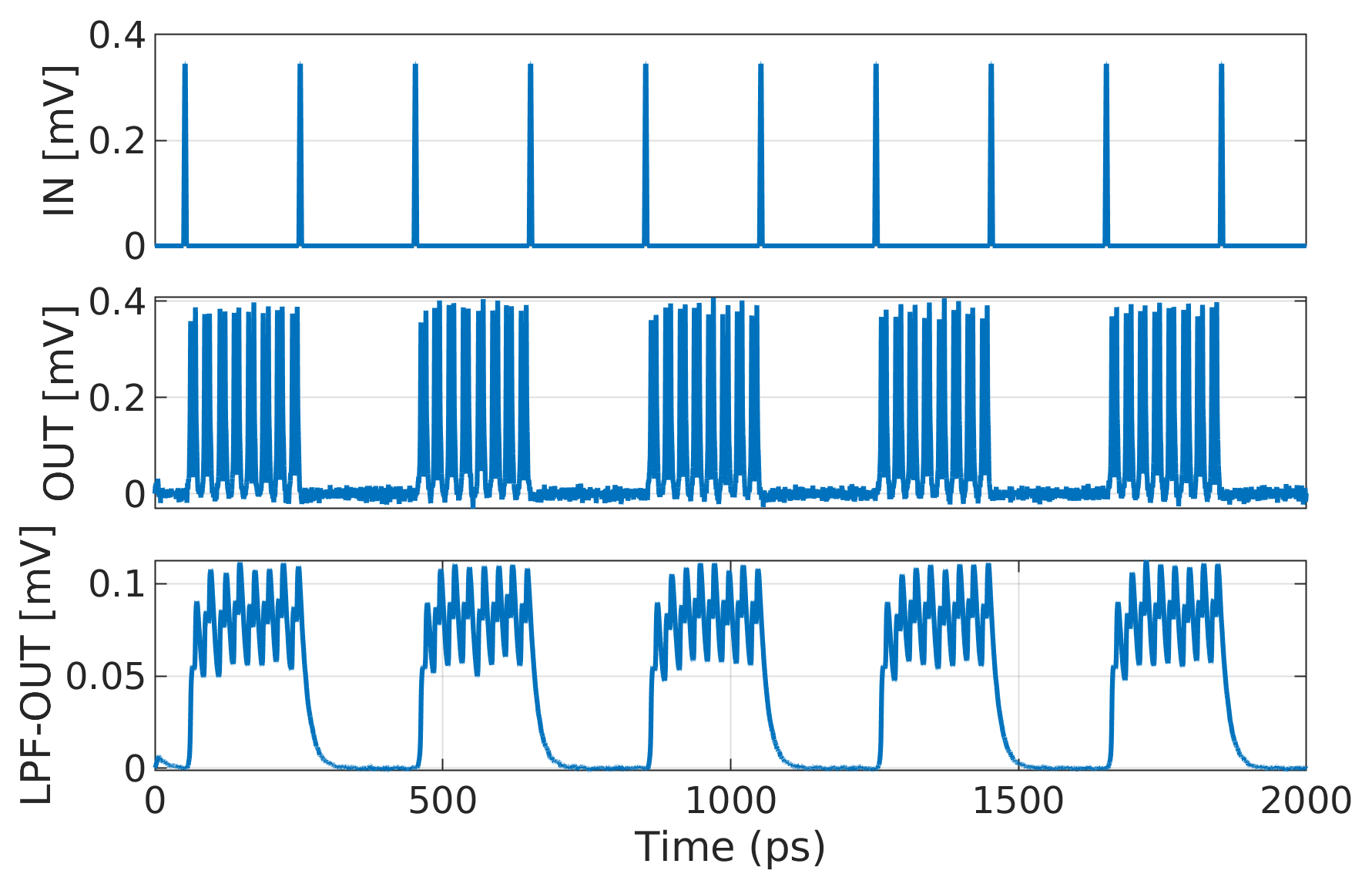}
    \centering
    \captionof{figure}{Simulation waveform of the SFQ/DC converter. After passing through a low pass filter (LPF), the DC signal level shows around 100$\mu$V amplitude.}
    \centering
    \label{fig:SFQDC_sim}
% \end{figure}

% \begin{table}[h]
\centering
\captionof{table}{Parameter values and margins of SFQ/DC converter} 
\label{table:SFQDC}
\resizebox{\textwidth}{!}{%
\begin{tabular}{|c|c|c|c|c|c|}
\hline
     Components & Values & Margins & Components & Values & Margins \\
     \hline
     J1& $112\mu A$ & $91\%$&
     J2& $104\mu A$ & $100\%$\\
     J3& $70\mu A$ & $100\%$&
     J4& $78\mu A$ & $100\%$\\
     J5& $70\mu A$ & $97\%$&
     J6& $80\mu A$ & $100\%$\\
     J7& $80\mu A$ & $100\%$&
     J8& $80\mu A$ & $100\%$\\
     L1 &  $0.5pH$ & $100\%$&
     L2 &  $3pH$ & $93\%$\\
     I1 & $45\mu A$ & $100\%$&
     I2 & $26\mu A$ & $100\%$\\
     I3 & $48\mu A$ & $100\%$&
     I4 & $22\mu A$ & $100\%$\\
     Rout & $50ohm$ & $100\%$&
     Bias  & $1mV$ & $87\%$ \\
\hline
\end{tabular}}
% \end{table}
\end{minipage}%
\end{figure}

\noindent
Fig.\ref{fig:SFQDC_sch} shows the SFQ/DC converter schematic. When an input pulse comes, it breaks the quiescent state of the cell and leads the output junction J5 to start oscillating. Another input pulse will then pull the cell back to its initial state. While in active mode, the SFQ/DC converter will keep pumping out current through the L1. The serial resistor $R_{out}$ forms a low pass filter with the off-chip wire inductance and the equivalent load of the oscilloscope, which is usually used to monitor the output of a chip. Fig.\ref{fig:SFQDC_sim} is the simulation waveform. The input signal is at the top. The middle plot is the observed voltage after L1, and the bottom is the signal at the oscilloscope. As we can see, the output state changes every time an input pulse comes. The component values are listed in Table.\ref{table:SFQDC}. The critical margin of the PTL driver is $87\%$ on the overall bias voltage.
\section{Results and Implementation}
\subsection{Random Pattern Generator}
\begin{figure}[!t]
\centering
\noindent\begin{minipage}{\linewidth}
% \begin{figure}[h]
    \includegraphics[width=0.9\textwidth]{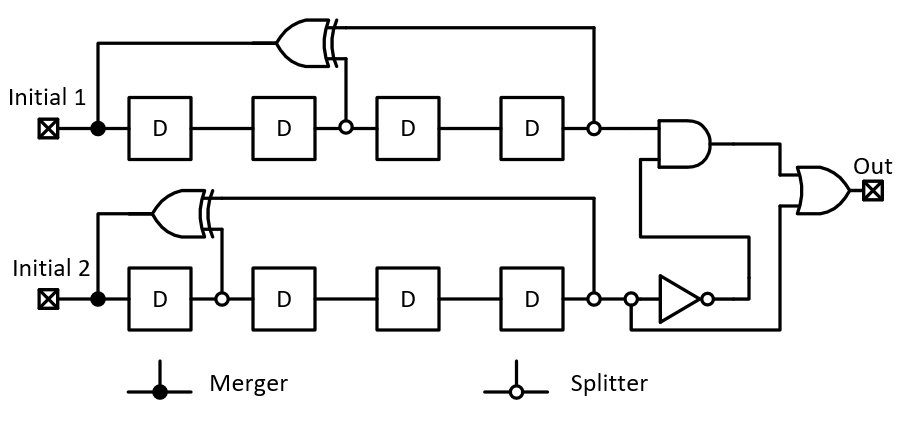}
    \centering
    \captionof{figure}{Schematic of the random pattern generator.}
    \centering
    \label{fig:circuit_sch}
% \end{figure}

% \begin{figure}[h]
    \includegraphics[width=1\textwidth]{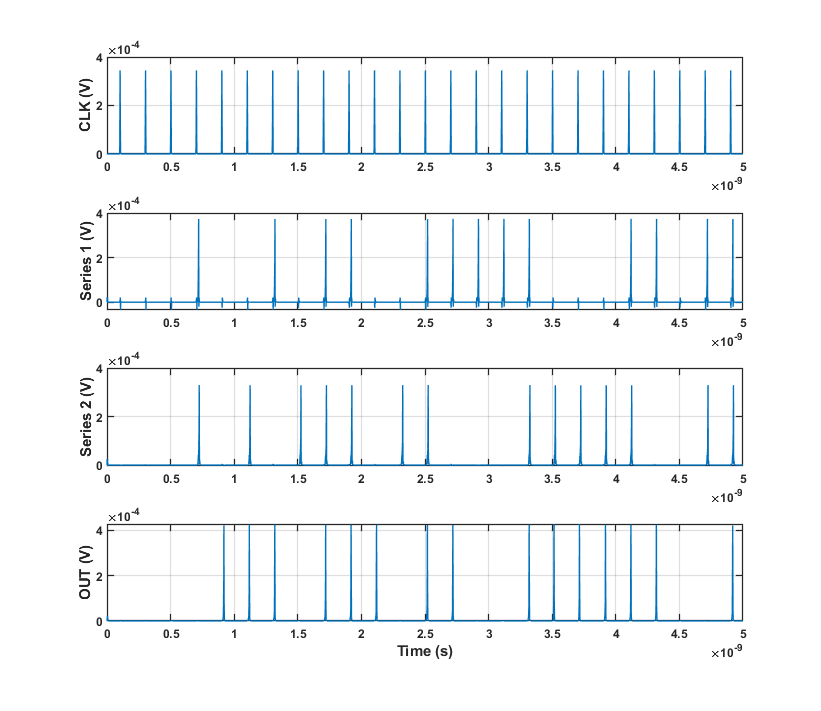}
    \centering
    \captionof{figure}{Simulation waveform of the random pattern generator.}
    \centering
    \label{fig:circuit_waveform}
% \end{figure}
\end{minipage}%
\end{figure}

\noindent
A random pattern generator has been implemented to evaluate the library cells, as depicted in the schematic shown in Fig.\ref{fig:circuit_sch}. Two DFF chains are formed, with signals tapped from various points in these chains using splitters. These signals are then directed to XOR gates. Subsequently, the outputs of the two XOR gates are combined with the initial signals, creating a feedback data loop. This configuration generates two random data series inputs to the subsequent stages: the AND gate, inverter, and OR gate. All cells in Fig.\ref{fig:circuit_sch} are synchronized with a clock signal, although the clock signal is not displayed to maintain diagram clarity and readability.

Fig.\ref{fig:circuit_waveform} presents the simulation waveform of this microsystem. The top waveform represents the clock signal distributed to all the cells. 'Series 1' and 'Series 2' denote the bit series at the outputs of the two DFF chains, while 'OUT' signifies the output of the final OR gate. This microsystem effectively demonstrates the correct functionality of the employed cells and showcases the system integration capability of the half-flux quantum standard cells.

\begin{figure}[!ht]
    \includegraphics[width=0.45\textwidth]{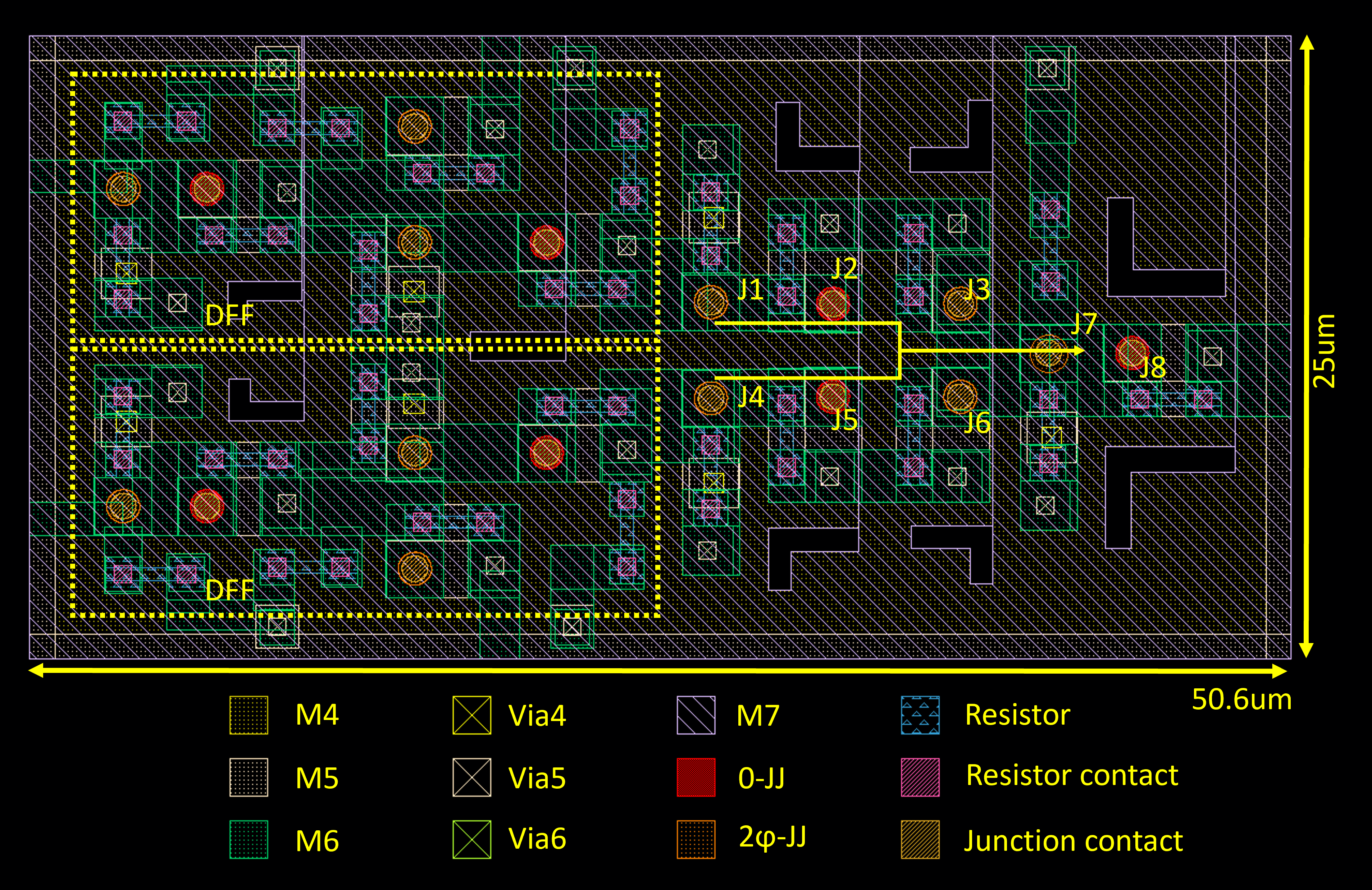}
    \centering
    \caption{An example layout of the OR gate.}
    \centering
    \label{fig:layout}
\end{figure}

The above explanation described the HFQ standard cell library design utilizing the $2\phi$-junction. Additionally, prototype layouts were created for each cell using a simulated technology based on the published MITLL SFQ5ee process. An extra layer, the $2\phi$-junction layer, was assumed to enable the implementation of the $2\phi$-junction. Fig.\ref{fig:layout} provides an illustration of the OR gate layout. The layout predominantly comprises four metal layers (M4 - M7), three metal vias, one resistor layer, two junction layers for 0-JJ and $2\phi$-JJ respectively, one resistor contact layer, and one junction contact layer.

In Fig.\ref{fig:layout}, it's evident that eliminating the inductor results in a more compact layout. Furthermore, as technology advances, the area could potentially be further reduced. Table.\ref{table:summary} provides a comprehensive list of all the cells implemented in this library, including the number of junctions (0-JJ and $2\phi$-JJ) and critical margins. It also compares estimated layout areas and bias current with a conventional RSFQ library we implemented. On average, the HFQ cells exhibit a 50.8\% reduction in area and a 61\% decrease in bias current compared to the conventional RSFQ library.

\begin{table*}[ht]
\centering
\caption{Summary of the $2\phi$-JJ based library cells} 
\label{table:summary}
\begin{tabular}{|c|c|c|c|c|c|c|}
\hline
     Cell names & Number of JJs & Critical margins & Areas & Bias current & Areas (RSFQ) & Bias current (RSFQ)\\
     \hline
     JTL & 4 & $71\%$ & $225 um^2$ & $90 \mu A$ & $625 um^2$ & $180 \mu A$ \\
     DFF & 5 & $68\%$ & $375 um^2$ & $63 \mu A$ & $625 um^2$ & $212 \mu A$ \\
     NDRO & 9 & $55\%$ & $625 um^2$ & $487 \mu A$ & $2500 um^2$ & $863 \mu A$ \\
     Merger & 8 & $72\%$ & $550 um^2$ & $174 \mu A$ & $625 um^2$ & $375 \mu A$ \\
     Splitter & 7 & $70\%$ & $494 um^2$ & $160 \mu A$ & $625 um^2$ & $309 \mu A$ \\
     OR & 18 & $37\%$ & $1263 um^2$ & $259 \mu A$ & $2500 um^2$ & $475 \mu A$ \\
     AND & 18 & $85\%$ & $1263 um^2$ & $220 \mu A$ & $2500 um^2$ & $530 \mu A$ \\
     XOR & 19 & $20\%$ & $1428 um^2$ & $290 \mu A$ & $2500 um^2$ & $435 \mu A$ \\
     %INV & 15 & $16\%$ & $1469 um^2$ & $324 \mu A$ & $2500 um^2$ & $546 \mu A$ \\
     PTL driver & 4 & $81\%$ & $429 um^2$ & $93 \mu A$ & $625 um^2$ & $265 \mu A$ \\
     PTL receiver & 4 & $81\%$ & $369 um^2$ & $107 \mu A$ & $625 um^2$ & $252 \mu A$ \\
     DC/SFQ converter & 3 & $98\%$ & $479 um^2$ & $48 \mu A$ & $1875 um^2$ & $450 \mu A$ \\
     SFQ/DC converter & 8 & $87\%$ & $876 um^2$ & $141 \mu A$ & $3125 um^2$ & $1025 \mu A$ \\
\hline
\end{tabular}
\end{table*}

\section{Conclusion}
\label{sec:conc}
\noindent
An HFQ standard cell library employing the $2\phi$-junction is demonstrated. The detailed design methodology encompasses schematic representations, component values, and their respective margins for each available block within the standard cell library. The library includes essential components such as inverters, AND, OR, XOR gates, JTLs, splitters, mergers, PTL drivers, PTL receivers, DFFs, DC/HFQ converters, and HFQ/DC converters.
Compared to conventional RSFQ cells, this new design necessitates less bias current, reduces reliance on inductors, enhances stability and scalability, and occupies about 55\% smaller area. These advancements position the HFQ logic family as a compelling contender for the next generation of VLSI circuits.

% use section* for acknowledgment
\section*{Acknowledgment}
This work was partly supported by the National Science Foundation (NSF) through the project Expedition: Discover (Design and Integration of Superconducting Computation for Ventures beyond Exascale Realization) under Grant 2124453.
% Can use something like this to put references on a page
% by themselves when using endfloat and the captionsoff option.
\ifCLASSOPTIONcaptionsoff
  \newpage
\fi

\end{document}